\documentclass[aps,pra,twocolumn,superscriptaddress,amsmath,amssymb,showpacs,notitlepage,nofootinbib]{revtex4-1}
\pdfoutput=1
\usepackage[T1]{fontenc}
\usepackage{graphicx}% Include figure files
\usepackage{dcolumn}% Align table columns on decimal point
\usepackage{bm}% bold math
\usepackage{hyperref}% add hypertext capabilities
\usepackage{tikz}
\usepackage{mathtools}
\usepackage{physics}
\usepackage{braket}
\usepackage{hyperref}
\usepackage{multirow}
\usepackage{stackengine}
\usepackage{array}
\usepackage{xcolor,color}
\usepackage{lipsum}
\usepackage{enumitem}
\usepackage{dblfloatfix} 
\usepackage[caption=false]{subfig}
\usepackage{dsfont}
\usepackage[normalem]{ulem}
\usepackage{algorithm}
\usepackage[noend]{algpseudocode}
\usepackage{float}

\begin{abstract}
One of the most promising applications of quantum networks is entanglement assisted sensing. The field of quantum metrology exploits quantum correlations to improve the precision bound for applications such as precision timekeeping, field sensing, and biological imaging. When measuring multiple spatially distributed parameters, current literature focuses on quantum entanglement in the discrete variable case—and quantum squeezing in the continuous variable case distributed amongst all of the sensors in a given network. However, it can be difficult to ensure all sensors pre-share entanglement of sufficiently high fidelity. This work probes the space between fully entangled and fully classical sensing networks by modeling a star network with probabilistic entanglement generation that is attempting to estimate the average of local parameters. The quantum Fisher information is used to determine which protocols best utilize entanglement as a resource for different network conditions. It is shown that without entanglement distillation there is a threshold fidelity below which classical sensing is preferable. For a network with a given number of sensors and links characterized by a certain initial fidelity and probability of success, this work outlines when and how to use entanglement, when to store it, and when it needs to be distilled.
\end{abstract}
\begin{document}

\preprint{APS/123-QED}

\title{Utilizing probabilistic entanglement between sensors in quantum networks}

\author{Emily A. Van Milligen}
\email{evanmilligen@arizona.edu}
 \affiliation{Department of Physics, The University of Arizona, 1630 East University Boulevard, Tucson, AZ 85721}
\author{Christos N. Gagatsos}%
\affiliation{Department of Electrical and Computer Engineering, The University of Arizona, 1630 East University Boulevard, Tucson, AZ 85721}
\affiliation{Wyant College of Optical Sciences, The University of Arizona, 1630 East University Boulevard, Tucson, AZ 85721}
\affiliation{Program in Applied Mathematics, The University of Arizona, Tucson, Arizona 85721, USA}
\author{Eneet Kaur}
\affiliation{Wyant College of Optical Sciences, The University of Arizona, 1630 East University Boulevard, Tucson, AZ 85721}
\affiliation{Cisco Quantum Lab, Los Angeles, USA}
\author{Don Towsley}
\affiliation{College of Computer Science, University of Massachusetts, Amherst, MA}
\author{Saikat Guha}%
\affiliation{Wyant College of Optical Sciences, The University of Arizona, 1630 East University Boulevard, Tucson, AZ 85721}
\affiliation{Department of Electrical and Computer Engineering, The University of Arizona, 1630 East University Boulevard, Tucson, AZ 85721}
\date{\today}
\maketitle
%\begin{abstract}%

%Octogons for sensors in images? Check the color looks good in black and white and printed
%argue that there are large T2 times compared to (k-1)\tau and L/c so there is negligable degradation
\section{\label{Introduction}Introduction}
Quantum networks are equipped to distribute entanglement to consumers who utilize it as a resource for a variety of applications. Some of the most promising applications are distributed computing, which will allow for scalable quantum computers \cite{7562346}, entanglement-assisted communications, which allow for the Holevo capacity limit to be surpassed \cite{9173940,PhysRevLett.126.250501}, and quantum cryptography, which allows for secure encryption and transmission protected by the rules of quantum mechanics \cite{Ekert1992,PhysRevLett.68.557}. Certain distributive sensing problems have also been shown to benefit from entanglement between sensors, as quantum correlations can help improve the  precision bound when estimating unknown parameters \cite{PhysRevLett.124.150502,PhysRevLett.129.180502,PhysRevLett.111.070403,PhysRevA.97.032329}. 

Current literature on quantum metrology includes results exploiting quantum entanglement in the discrete variable case and quantum squeezing in the continuous variable case between all the sensors in a given network \cite{giovannetti_advances_2011,pirandola_advances_2018,zhang_distributed_2021}. This study focuses on the discrete variable case, where it has already been shown that entanglement between sensors can provide significant improvement when measuring functions of parameters, such as when taking the average \cite{PhysRevLett.120.080501,Bringewatt_2021, rubio_quantum_2020}. However, ensuring all sensors pre-share entanglement can be difficult for a number of reasons. Creating and distributing entangled states between sensors can be probabilistic, so multiple attempts may be required before all sensors share an entangled state. This work analyzes such a case, where some subset of sensors pre-share entanglement while others sense with local probe states, and details some protocols to minimize the precision bound when sensing an unknown global parameter.

This paper has four major contributions. Firstly, we model entangled pair generation as a probabilistic process and introduce time multiplexed quantum memories that can connect pairs that are generated at different times. With this model in place, we introduce three sensing protocols: an \textit{Immediate Sensing} protocol, a \textit{Fixed Time Multiplexing Block Length} (F-TMBL) protocol, and a \textit{Variable Time Multiplexing Block Length} (V-TMBL) protocol. Secondly, we explore how the number of sensors sharing an entangled state should be chosen based on the quality of the entanglement. We then discuss how distillation can be incorporated into these protocols and show when it is beneficial to wait for more entangled pairs in order to distill higher quality entanglement. Lastly, we give the optimal measurements for our protocols and compare against the performance of local measurements. 

We will define our paper's notation and formally set up our sensing problem in Section II and explain the concept of quantum Fisher information (QFI) as a measure evaluating the effectiveness of our sensing protocols. In Section III, we introduce our sensing protocols and evaluate their performance when entanglement generation results in pure states with perfect fidelity. In Section IV, we loosen this condition and assume that entanglement generation results in mixed states. This section discusses how the form of the probe state should be chosen based on the initial quality of the entangled pairs. In Section V, we show how distillation can be incorporated into these protocols. In Section VI, we discuss the optimal measurements to perform on the sensors to extract the unknown parameter and compare against the results of a strategy strictly limited to local measurements. Finally, we will provide some concluding statements in Section VII and suggest future directions for this work.

\section{\label{Problem Setup}Problem Setup}
  \begin{figure}
    \begin{centering}
        \subfloat[Quantum Sensing Network]{\includegraphics[width=\linewidth]{ 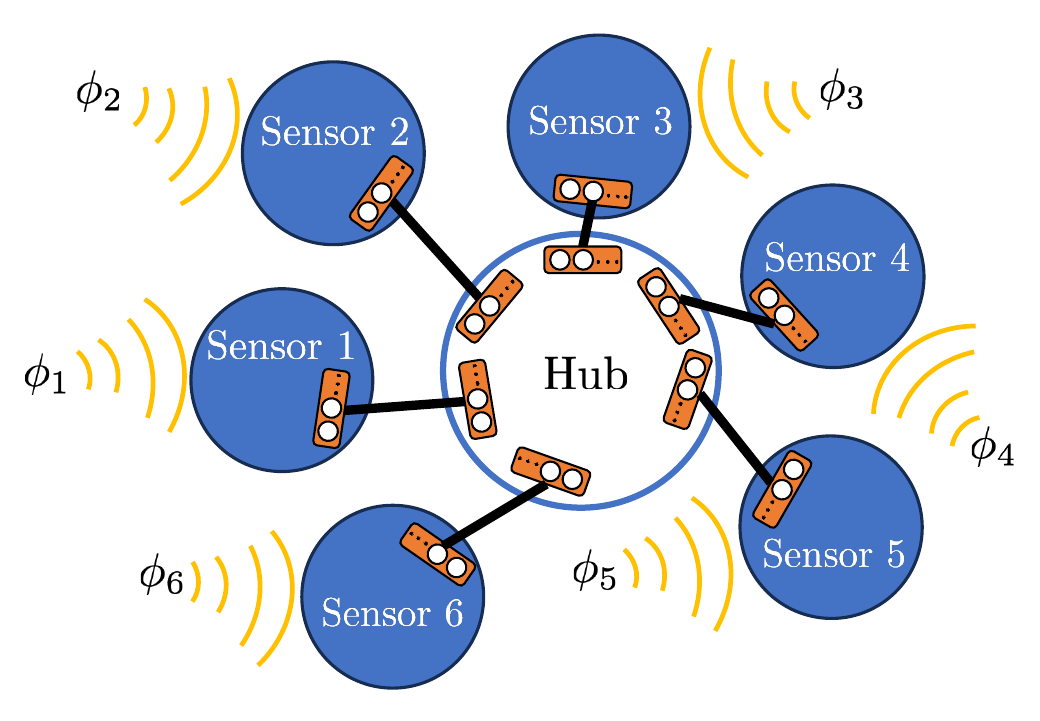} \label{Quantum Sensing Network}} \hfill
        \subfloat[Network Snapshot] {\includegraphics[width=\linewidth]{ 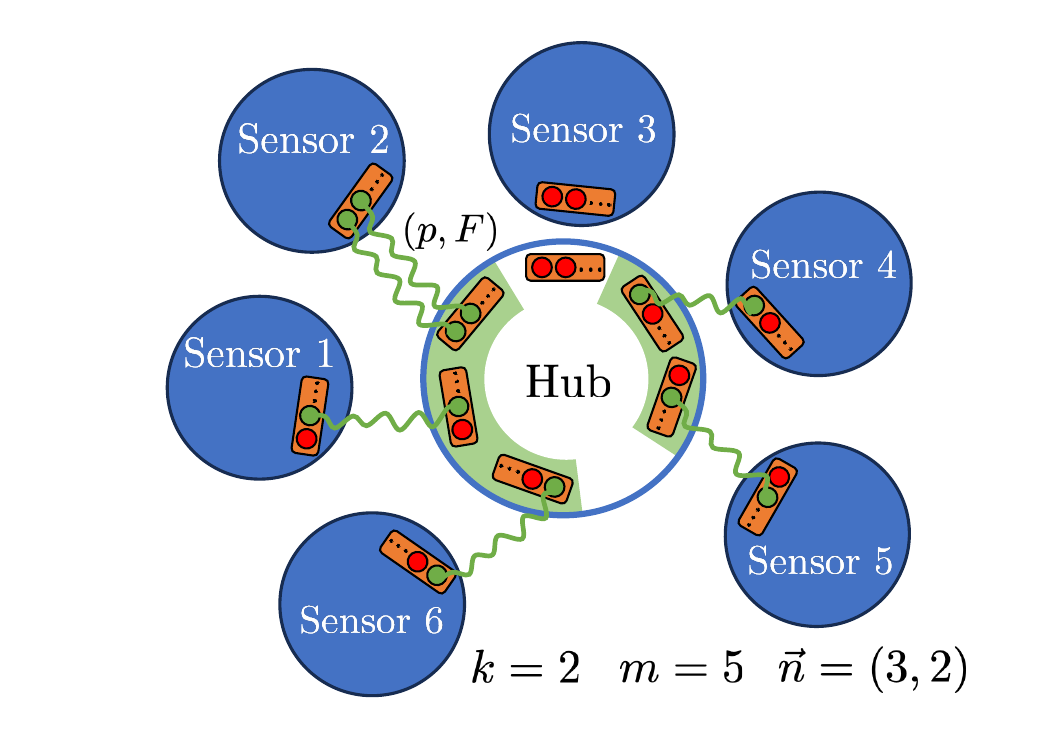}\label{Entanglement Snapshot}}
    \end{centering}
    \caption{$S=6$. (a) The $i$-th sensor is modulated by local parameter $\phi_i$. Each sensor has a queue of qubits to probe with, depicted as a box containing circles. They also have optical and classical communication channels connecting them with the hub, depicted by black lines, over which entangled pairs can be generated. The hub node has ensembles of qubits and is also equipped with dynamic circuits to perform GHZ projective measurements (not depicted). (b) Successful links are shown by green wavy lines connecting green qubits and failed generation attempts are shown by red qubits with no connecting lines. Links succeed with probability $p$ and have initial fidelity $F$. This snapshot shows the Fixed Time Multiplexing Block Length protocol for $k=2$. The number of successful connections are $m=5$ and the hub performed a 3-GHZ projective measurement and a 2-GHZ projective measurement, i.e. $\vec{n}=(3,2)$.}
    \label{fig:1}
\end{figure}
Following the notation found in \cite{PhysRevLett.120.080501}, we will consider the estimation problem where there are $S$ sensors and $S$ unknown local parameters denoted $\vec \phi=(\phi_1,\phi_2,...,\phi_S)^{T} \in \mathbb{R}^S$ . Let the parameters be imprinted on the system via a local unitary evolution such that the $j$-th sensor is acted on by $U_j=e^{-i\hat{H}_j\phi_j}$ where $\hat{H}_j$ acts non-trivially only on the Hilbert space of the $j$-th sensor. Therefore $\hat{H}_j$ can be expressed as $\hat{H}_j=(\mathds{1}\otimes...\otimes \hat{h}_j \otimes..\otimes\mathds{1})$ where $\hat{h}_j$ is the nontrivial action on the $j$-th qubit. The resulting state of the system of sensors is $\rho_{\vec \phi}=U(\vec \phi)\rho U(\vec \phi)^{\dagger}$ where $U(\vec \phi) = e^{-i\vec \phi^{\dagger}\hat{H}}$ and $\rho$ is the probe state.  Operators $\hat{H} = (\hat{H_1},\hat{H_2},...\hat{H_S})$ are Hermitian and  mutually commute. This general problem formulation can be used to model many quantum sensing problems, including phase shift estimation for quantum imaging \cite{aasi_enhanced_2013}, field mapping \cite{PhysRevLett.116.030801} and coordinating networks of clocks \cite{komar_quantum_2014}.

For certain problems, entanglement between sensors may not be beneficial \cite{PhysRevLett.120.080501}. We chose a linear function of these parameters $\theta (\vec \phi)$ to estimate, instead of sensing the $S$ parameters separately. We choose $\theta(\vec \phi)$ be the average value of $\vec \phi$, i.e.: $\theta(\vec \phi)=\frac{1}{S}(\phi_1+\phi_2+...+\phi_S)$, so that the largest improvement when using entanglement can be attained \cite{Bringewatt_2021, rubio_quantum_2020}. For cleaner notation henceforth, the parenthetical will be dropped and it will be written as $\theta$.

\subsection{Probe State Construction}
When sensing \textit{global} properties of the system that are a function of the local parameters, such as in the case of $\theta$, the optimal protocol often requires use of entangled probe states. For certain sensing problems, it has already been shown that a GHZ state, a type of maximally entangled state, is the optimal probe \cite{PhysRevA.54.R4649,PhysRevLett.79.3865}. There are a number of ways to prepare such a state. In this paper, we will let the sensors be capable of forming Bell pairs with some central hub node that contains at least $S$ arrays of quantum memories, one for each sensor, as shown in Fig. \ref{fig:1}. These Bell pairs, which may also be referred to as links,  can be written as
\begin{align}
\frac{\ket{0}_{A}\ket{0}_{B}+\ket{1}_{A}\ket{1}_{B}}{\sqrt{2}},
\end{align}
where $A$ denotes a qubit held at the sensor and $B$ denotes a qubit held at the hub. (Note: $\ket{0}$ and $\ket{1}$ are qubit states in the computational basis.) Each sensor has probability $p$ of successfully generating an entangled link with the hub. The value of $p$ is dependent on the actual method of entanglement generation and the physical hardware. For example, we could consider a swap in the middle architecture where atomic qubits are held at the sensors and the hub emits entangled photonic qubits which are then are mixed via a balanced beamsplitter and finally detected by photon number resolving detectors \cite{Barrett_2005}. In this scenario, $p$ would be limited by the probability of detecting a specific click pattern that heralds success as well as by the detector inefficiencies such as dark click probabilities, dead times, etc. 

The hub node is capable of performing any \textit{n}-GHZ projective measurements on the qubits it has stored, where $n \in [2,...,S]$.  If the $n$-GHZ projective measurements succeed, the result is a $n$-GHZ state held between the sensing qubits described by
\begin{equation} \label{eq:GHZdef}
|\Gamma(n)\rangle=\frac{|{0}\rangle^{\otimes n}+|{1}\rangle^{\otimes n}}{\sqrt{2}},
\end{equation}
  up to a unitary. Note that if $n=2$, this is just a Bell state. For the purposes of this paper, we will assume that these measurements are deterministic as well as perfect, i.e. they always succeed and do not introduce any additional gate noise. In certain cases, we can push the GHZ projection noise to the state preparation. There are of course other ways to prepare/distribute GHZ states to sensors, some of which are discussed in \cite{avis_analysis_2023}.

Assume that after an initial latency, discussed in detail in Appendix \ref{Appendix:Latency}, each sensor simultaneously attempts to generate entanglement with the hub node every $\tau$ seconds. Any GHZ projective measurements performed on the resulting successful states will be assumed to be instantaneous. Meanwhile, all sensors will simultaneously probe the system every $\tau$ seconds. It is not necessary for these time scales to be the same, however since the asymptotic behavior is being evaluated by this paper it is a justified assumption. Therefore, we can think of time as being slotted. The system will be sensed $N$ times, such that the total sensing time will be $N\tau$ in the limit $N\xrightarrow{} \infty$. Then the overall probe state of the system accounting for all these trials can be written as $\rho=\rho(1)\otimes\rho(2)\otimes...\otimes\rho(N)$, where $\rho(t)$ is the $t$-th probe state used to measure $\theta$. Another way to say this is $\rho(t)$ is the probe state used in the $t$-th timeslot.

A snapshot of the network after a generation attempt can be characterized by the number of successful Bell pairs generated between the sensors and the hub.
Assuming that initially no sensors are entangled with the
hub, the number of sensors entangled with the hub after a generation attempt is a random variable $M$ given by the binomial
distribution
\begin{equation}\label{binomial prob}
\text{Pr}(M=m)={{S}\choose{m}}p^m(1-p)^{S-m},
\end{equation}
 where $0\leq m \leq S$. After pairwise entanglement is generated, a GHZ projective measurements may be performed depending on the protocol.
If there are $m$ successful links in a particular instance of the network, a $n$-GHZ-projective measurement can made on the snapshot, where $2 \leq n \leq m$ depending on what the protocol dictates. It will be assumed that all sensors not involved in this projective measurement will prepare some known local state to probe with. The actual local state should be chosen based on the form of the unitary $U(\vec \phi)$. For this paper, all local probe states will be in the $\ket{+}$ state, defined as
\begin{equation}\label{eq:plus}
\ket{+}=\frac{|{0}\rangle+|{1}\rangle}{\sqrt{2}}.
\end{equation}
The preparation of these local states will be assumed to result in $\ket{+}$ states of perfect fidelity. Initially, unit fidelity Bell pairs and GHZ projective states will be modelled to form an initial understanding of the problem. However, this condition will be relaxed in Sec. \ref{Non-Unit Fidelity Bell Pairs}, as it is not a realistic representation of a physical system. The Bell pairs will thereafter be modelled as Werner states with initial fidelity $F$.

\subsection{Quantum Fisher Information}
After probing the system with $\rho$, a set of  positive-operator valued measures (POVMs) can be applied. An estimate of $\theta$ can then be calculated based on the outcome. This work focuses primarily on finding the optimal probe state, independent of the measurement scheme. For this reason, we use the quantum Fisher information (QFI) as the metric of interest. The QFI provides a lower bound of the variance of the estimate of $\theta$ based solely off the form of the intital probe state.

The Quantum Cramer Rao Bound (QCRB) can be expressed as
\begin{equation}
\text{Var}(\theta)\geq \frac{1}{\sum_{t=1}^{N}\mathcal{F}_\theta(\rho (t))},
\end{equation}
where $\mathcal{F}_\theta(\rho (t))$ is the Fisher Information corresponding with probe state of the $t$-th trial. If the probe state is the same for all time steps, one finds the bound
\begin{equation}{\label{QCRB-single}}
\text{Var}(\theta)\geq \frac{1}{N\mathcal{F}_\theta}.
\end{equation}
The quantity of interest for the remainder of this paper will be the average QFI which can be calculated as
\begin{equation}
\bar{\mathcal{F}_\theta}=\frac{1}{N}\sum_{t=1}^{N}\mathcal{F}_\theta(\rho (t)),
\end{equation}
when $N\xrightarrow{}\infty$. It will be shown  how it varies with $p,\ S,$ and $F$. From there, different protocols will be compared to see what maximizes this quantity.

\section{Unit Fidelity Bell Pairs\label{Unit Fidelity Bell Pairs}}
Initially, it will be assumed that entanglement generation, though probabilistic, results in perfect Bell pairs held between the sensors and the hub. Additionally, it will be assumed that no decoherence occurs while the qubits are stored in the quantum memories. First, the QFI for a given network snapshot will be calculated and then three different protocols for sensing will be explored so that the average QFI of the system over time can be maximized.

\subsection{Network Snapshot Expression}
If $m$ sensors share entanglement with the hub after a entanglement generation attempt, then a $n$-GHZ projective measurement can be performed where $2\leq n\leq m$.
The resulting probe state can be written as
\begin{equation} \label{eq:General State}
|\psi_n\rangle=\left(\frac{|{0}\rangle^{\otimes n}+|{1}\rangle^{\otimes n}}{\sqrt{2}}\right) \otimes \left(\frac{|{0}\rangle+|{1}\rangle}{\sqrt{2}}\right)^{\otimes ({S}-n)}, 
\end{equation}
where  $\rho_n=\ket{\psi_n}\bra{\psi_n}$. This state is tensor product between a $n$-GHZ state (the first term) and $S-n$ local superposition states, (the second term). For the purposes of this paper, we will also define $\ket{\psi_0}=\ket{\psi_1}=\ket{+}^{\otimes S}$. When the subscript takes these values, we are not implying a $n=0$ or a $n=1$ - GHZ projective measurement is performed. Instead, we will adopt this notation to indicate that the probe state is a product state in order to keep calculations consistent.

The computational basis we use consists of the eigenvectors of the Hamiltonian so that if the $j$-th qubit is in the state $\ket{0}_j$ then $\hat{h}_j\ket{0}_j=\lambda_{0}\ket{0}_j$ and $\hat{h}_j\ket{1}_j=\lambda_{1}\ket{1}_j$. For the purposes of this paper, we will set $\hat{h}_1=\hat{h}_2=\cdots=\hat{h}_S$ so that all of the eigenvalues are the same. The Fisher information associated with a certain snapshot/probe state is found to be
\begin{equation}\label{QFIsimplesnap}
\mathcal{F}_\theta(\rho_n)=\frac{(\Delta\lambda)^2 }{S^2}(S-n+n^2),
\end{equation}
where $\Delta\lambda=\lambda_{1}-\lambda_{0}$. For the figures in this paper, we set $(\Delta\lambda)^2=1$ as it is a constant scaling factor.
In this form, it is easy to distinguish the $S-n$ term coming from the $S-n$ local probes and the $n^2$ term coming from the $n$ sensors that share entanglement. Note that when less than two sensors connect to the hub, the hub is unable to perform a GHZ projective measurement. In this case, we find $\mathcal{F}_\theta(\rho_n)=\frac{(\Delta\lambda)^2}{S}$, which agrees with what we get from the above equation when setting $n=0$ or $n=1$. However, if all of the sensors pre-share entanglement, i.e., $n=S$, we get $\mathcal{F}_\theta(\rho_n)=(\Delta\lambda)^2$, which shows the proper scaling $S^{-1}$ improvement attained by previous results. The derivation of a more generalized version of this equation can be found in Appendix~\ref{QFI_Calc}.

This equation can be extended to account for probe states consisting of product states of multiple GHZ states. If there are $g$ different GHZ states, such that the $\nu$-th GHZ is a $n_{\nu}$-GHZ state, where $\nu\in [1,g]$ we get 
\begin{equation}
\label{unit-fidelity qfi}
\mathcal{F}_\theta(\rho_{\vec{n}})=\frac{(\Delta\lambda)^2 }{S^2}\left(S+\sum_{\nu=1}^g\left(n_{\nu}^2-n_{\nu}\right)\right),
\end{equation}
where $\vec{n}=(n_1,n_2,\cdots,n_g)$ and $\sum_{\nu=1}^gn_{\nu}\leq m$. An example of this is shown in Fig. \ref{fig:1} (b), which shows the case for $m=5$, $\vec{n}=(3,2)$.
This expression shows the trade off in the QFI between having fewer entangled sensors in a larger entangled components as opposed to having many smaller entangled components. The optimal partitioning of successes that maximizes the QFI occurs when a $m$-GHZ projective measurement is performed, which entangles all the sensors that successfully connected to the hub. This optimization problem grows more complex however when $F<1$, which will be discussed in Section \ref{partitioning}.
\begin{figure}
\includegraphics[width=0.45\textwidth]{ 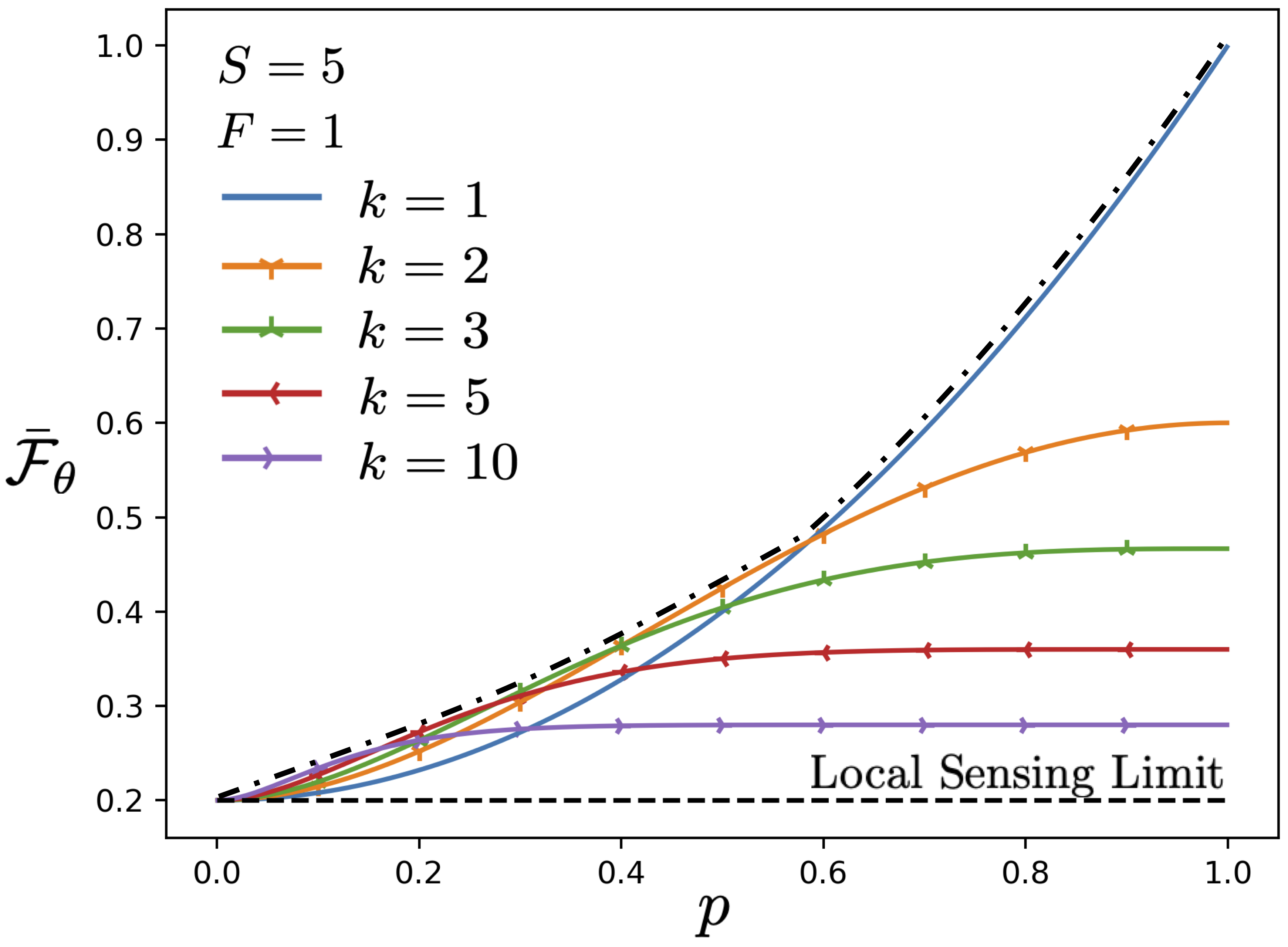}
\caption{\label{fig:2} Average QFI attained by the F-TMBL protocol vs $p$ given $S=5$ and $F=1$ for a variety of $k$ values. When $p<2-\sqrt{2}$, there is an optimal $k>1$ that maximizes the average QFI. The dashed line at $1/5$ shows the QFI associated with a probe state where all the sensors locally prepare $\ket{+}$. ($(\Delta\lambda)^2=1$)}
\end{figure}
\subsection{Sensing Protocols\label{Sensing Protocols}}
Now that the network architecture and the sensing problem has been established, three different network protocols are introduced. These protocols are: \textit{Immediate Sensing}, \textit{Fixed Time Multiplexing Block Length} (F-TMBL), and \textit{Variable Time Multiplexing Block Length} (V-TMBL). The average QFI associated with each will be compared in order to gauge how these protocols compare across different network conditions.
\begin{figure}
\begin{centering}
        \subfloat[]{\includegraphics[width=\linewidth]{ 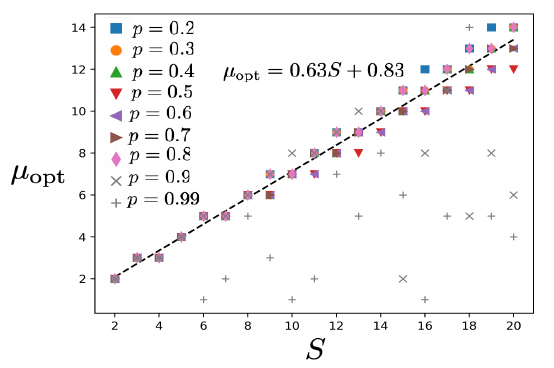} } \hfill
        \subfloat[] {\includegraphics[width=\linewidth]{ 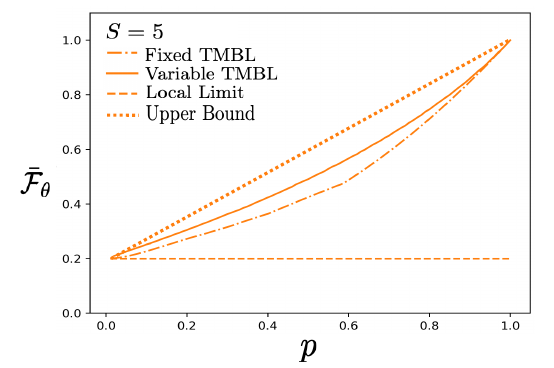}}
        \hfill
         \subfloat[] {\includegraphics[width=\linewidth]{ 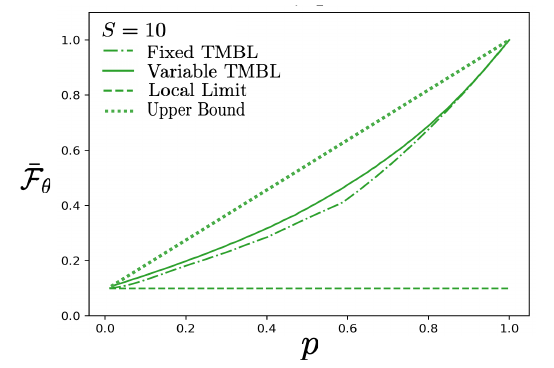}}
    \end{centering}

\caption{\label{fig:3} (a) $\mu_{\text{opt}}$ vs $S$ for a variety of $p$ values. A trend-line and its equation are shown for the data when $p<0.9$. For high $p$ values, the relationship between $\mu_{\text{opt}}$ and $S$ is less clear. The average QFI for both the F-TMBL and the V-TMBL is plotted against $p$ for the case of $S=5$ in (b) and $S=10$ in (c). For each $p$ value the optimal value of $k$ and $\mu$ are used. The V-TMBL outperforms the F-TMBL protocol, however this advantage diminishes as $S$ increases in size. ($(\Delta\lambda)^2=1$)}
\end{figure}
\subsubsection{\label{Immediate Sensing Protocol}Immediate Sensing Protocol}
The Immediate Sensing protocol requires all the sensors to attempt to generate entanglement with the hub once before the hub performs the GHZ projective measurements on the successful links. For the case where $F=1$, the hub will perform a $m$-GHZ projective measurement entangling all the successfully connected sensors. The probe state resulting from this protocol is then used to measure the unknown parameter. If only a single sensor forms a Bell state with the hub, the entanglement is discarded. 

The average QFI for this protocol comes directly from combining Eqs. (\ref{binomial prob}) (\ref{QFIsimplesnap}) and the known expectation values of the binomial distribution. The average QFI is
\begin{align}
\bar{\mathcal{F}_\theta}&=\sum_{m=0}^S{{S}\choose{m}}p^m(1-p)^{S-m}\mathcal{F}_\theta(\rho_m)\nonumber\\
&=\frac{(\Delta\lambda)^2}{S^2}\left(S-\mathbb{E}[m]+\mathbb{E}[m^2]\right)\nonumber\\
&=\frac{(\Delta\lambda)^2 }{S}\big[1+(S-1)p^2\big].
\end{align}
 The initial latency associated with this protocol is $2L/c$. 

\subsubsection{Fixed Time Multiplexing Block Length Protocol}
 The Fixed Time Multiplexing Block Length (F-TMBL) protocol divides time into blocks each consisting of $k$ timesteps, where $1\leq k << N$. For each of these timesteps, all sensors attempt to generate entanglement, saving any successes until the end of the block. If there are multiple links formed between the sensor and hub, additional specifications must be made. Since initially it is assumed that the links are pure states, only the most recently generated link is kept. At the end of the $k$-th timestep, assuming more than one sensor has connected to the hub, GHZ projective measurements are made by the hub and and the entangled probe state is used to sense. An example snapshot of this protocol is shown in Fig. \ref{fig:1} (b). For the case where $F=1$, the hub always performs the maximal GHZ projecive measurement possible. During the first $k-1$ time steps in the block, all the sensors will measure the desired parameters without entanglement assistance. If only a single sensor forms a Bell state with the hub at the end of the $k$ timesteps, the entanglement is discarded.

The multiplexing block length $k$ can be optimized for each value of $p$. It should be noted that the Immediate Sensing protocol corresponds to the case of time multiplexing block length $k=1$, as the sensors only attempt entanglement generation once before measuring the unknown parameter. 

The average QFI for this protocol is
\begin{align}
\bar{\mathcal{F}_\theta}&=\frac{k-1}{k}\mathcal{F}_\theta(\rho_0)+\frac{1}{k}\sum_{m=0}^S\text{Pr}(M=m)_k\mathcal{F}_\theta(\rho_m),
%&=\frac{(\Delta\lambda)^2}{k S^2}\nonumber\\
%&\left(S(k-1)+\sum_{m=0}^S\text{Pr}(M=m)_k(S-m+m^2) \right)
\end{align}
 where $\text{Pr}(M=m)_k$ is given as
 \begin{align}
\text{Pr}(M=m)_k={S \choose m}(1-(1-p)^k)^m((1-p)^k)^{S-m}.
\end{align}
This is just the binomial distribution shown in Eq. \ref{binomial prob} where $p\rightarrow1-(1-p)^k$ describes the probability of at least one success in $k$ rounds. Similarly to before, we can use the known expectation values for the binomial distribution to  rewrite the expression for QFI as
\begin{align}
\bar{\mathcal{F}_\theta}=\frac{(\Delta\lambda)^2}{S}\big[1+(S-1)\frac{(1-(1-p)^k)^2}{k}\big].
\end{align}

The average QFI vs $p$ is plotted in Fig. \ref{fig:2} for $k\in \{1,2,3,5,10\}$. A lower bound at $S^{-2}$ corresponds to each sensor using local probe states that are not entangled. As $p\xrightarrow{}0$, all $k$ protocols converge to this value, since no sensors are able to connect to the hub and therefore the hub is unable to generate entangled states to sense with. When $k=1$, the QFI goes to 1 as $p$ increases. In the case of $p=1$, all sensors are able to connect to the hub and therefore can share an $S$-GHZ state to probe with, recovering the expected $S$ improvement in scaling. When $p$ is large, the QFI decreases with $k$. This is because each link is generating more than one Bell pair and all but one are discarded; hence at most one GHZ state is generated per block.  Fig. \ref{fig:2} depicts the case of $S=5$, however the optimal $k$, $k_{\text{opt}}$, is independent of $S$. In fact, the optimal value of $k$ only depends on $p$. When $F=1$, the Fixed F-TMBL protocol outperforms the Immediate Sensing protocol when $p<2-\sqrt{2}$.  The initial latency associated with this protocol is $2L/c+k$.

\subsubsection{Variable Time Multiplexing Block Length Protocol}
 The Variable Time Multiplexing Block Length (V-TMBL) protocol attempts entanglement generation until there are at least $\mu$ entangled links. If there are multiple links formed between a sensor and hub, only the most recently generated link is kept. After saving all entangled links until this criteria is met the maximum GHZ projective measurement is performed and the entangled probe state is then used to sense. During intermediate timesteps, all the sensors will measure the desired parameter without entanglement assistance.
 
 The average QFI for this protocol is
\begin{align}\label{VarQFI}
\bar{\mathcal{F}_{\theta}}=\sum_{t=1}^{\infty}\sum_{m=\mu}^S\text{Pr}(T=t,M=m)\left[\frac{(t-1)\mathcal{F}(\rho_0)}{t}+\frac{\mathcal{F}(\rho_m)}{t}\right],
\end{align}
where $T$ is a random variable describing the first timestep that $m\geq\mu$ sensors are entangled.

The joint probability probability $T = t$ and $M = m$ is given by

\begin{widetext}
\begin{align}
   \text{Pr}(T=t,M=m)= 
    \begin{cases}
      {S\choose m}(1-p)^{S-m}p^m, & t=1\\
      {S\choose m}(1-p)^{(S-m)t}\sum_{j=m-\mu+1}^m{m\choose j}\left(1-(1-p)^{t-1}\right)^{m-j}p^j(1-p)^{j(t-1)}, & t>1
    \end{cases}    
\end{align}
\end{widetext}
%\begin{align}\label{Varprob}
%\text{Pr}(T=t,M=m)=
%\sum_{\vec{\ell}}\frac{S!}{(S-m)!\ell_1!\cdots \ell_{t}!}(1-p)^{(S-m)t} %\prod_{i\in [1,\cdots,t]}\left((1-p)^{i-1}p\right)^{\ell_i}\nonumber\\
%\text{where } \vec{\ell}=(\ell_1\cdots \ell_{t}),\  \ell_{t}\in[m-%\mu+1,m],\  \sum_{i=1}^{t}\ell_i=m,
%\end{align}

where $\mu\leq m\leq S$. The probability for $t=1$ is straightforward, while the expression for $t>1$ can be derived by considering first the probability that $j<\mu$ links have succeeded on the $t-1$ timestep, and that the remaining $m-j$ succeed on the $t$-th timestep. For the case that $\mu=0$ and $\mu=1$, we recover the Immediate Sensing protocol again.

For given $S$ and $p$, there exists an optimal value for $\mu$, denoted $\mu_{\text{opt}}$. Values of $\mu_{\text{opt}}$ were found for a variety of $(S,p)$ pairs via Monte Carlo simulation corroborated by truncated evaluations of Eq. (\ref{VarQFI})  and are displayed in Fig. \ref{fig:3} (a). The value of $p$ does not appear to have a large impact on the value of $\mu_{\text{opt}}$ except for when $p$ is around 0.9 or higher. As $p$ approaches 1, there is a high probability that all links succeed each timestep. In this case, the V-TMBL for all values of the $\mu$ converge to that of the that of the Immediate Sensing protocol. A trend line is shown in Fig. \ref{fig:3}(a) calculated from the data when $p<0.9$. In this regime there appears to be a linear relationship between $\mu_{\text{opt}}$ and $S$.

The V-TMBL protocol with $\mu_{\text{opt}}$ and the F-TMBL protocols with $k_{\text{opt}}$ are compared in Fig. \ref{fig:3} (b) and (c) for two different values of $S$. An overall upper bound on the system is included in these diagrams, for the optimistic case that all links always succeed simultaneously with probability $p$. In this case,
\begin{align}
    \mathcal{F}_{\text{Bound}}=\frac{S-1}{S}p+\frac{1}{S}.
\end{align} 
The V-TMBL protocol performs better than the F-TMBL protocol, however its advantage diminishes as $S$ increases. The latency associated with this protocol is variable. However, an expression of the average latency and the average QFI can be found in Appendix \ref{Appendix:VTMBL}.

\section{Werner states\label{Non-Unit Fidelity Bell Pairs}}
In this section, we relax the assumption that the initial entangled pairs generated between the sensors and the hub are pure Bell states. Instead, their states will be represented as Werner states. Define
\begin{align*}
    \Phi_{AB}=\frac{\ket{00}_{AB}+\ket{11}_{AB}}{\sqrt{2}}\frac{_{AB}\bra{00}+_{AB}\bra{11}}{\sqrt{2}},
\end{align*}
to be a Bell pair with qubit $A$ held at the sensor and qubit $B$ held at the hub. Then a Werner state can be defined as
\begin{equation}
\beta_{AB} = x \Phi_{AB} + \frac{(1-x)}{4}\mathbb{I}_{A}\otimes\mathbb{I}_B,
\end{equation}
where  $x$ refers to the Werner parameter defined as $x=\frac{4F-1}{3}$ where $F$ is the fidelity. We chose to model the links as Werner states because any two qubit states with fidelity $F$ can be ``twirled" into Werner states \cite{PhysRevA.40.4277}. Doing this operation does not affect the fidelity, but mixes the contributions of the other Bell basis states. Werner states therefore can be considered to provide the worst case scenario in calculations, as presumably this twirling operation could be skipped.

\subsection{Network Snapshot Expression} 
When $n$ pairs of Werner states are prepared,  one qubit from each pair can be measured off at the hub node to prepare the following state: 
\begin{equation}
\sigma_{n}=\bra{\Gamma(n)}_{B_1\cdots B_{n}}\bigotimes_{i=1}^{n}\beta_{A_iB_i}\ket{\Gamma(n)}_{B_1\cdots B_{n}},\label{eq:GHZ-measurement}
\end{equation}
where $\ket{\Gamma(n)}$ is defined in (\ref{eq:GHZdef}).

This measurement diagonalizes the state in the $n$-GHZ basis.
The coefficients making up the different components of $\sigma_{n}$, were calculated in \cite{kaur2023entanglement} \cite{10313684}, and are included in Appendix \ref{Appendix:GHZcoeff}. The two that matter for the QFI are $\bra{\Gamma(n)}\sigma_{n}\ket{\Gamma(n)}$ and $\bra{\Gamma(n)}Z\sigma_{n}Z\ket{\Gamma(n)}$. The reason the rest are not important is because the projection of the state onto the GHZ basis vector $\Psi\not\in\{\ket{\Gamma(n)},Z\ket{\Gamma(n)}\}$ cancels out the contribution from the projection onto $\ket{\Psi'}=Z\ket{\Psi}$. For a further explanation, see the full derivations for the QFI found in Appendix \ref{QFI_Calc}.

%It is of note that the coefficients for states that vary only by a $Z$ gate are the same in all cases except for $\op{\Gamma(m)}$ and $Z_1\op{\Gamma(m)}Z_1$. This can be used later to simplify the QFI calculation.

 The entire probe state may consist of $g$ different GHZ diagonal states -- corresponding to the $g-$disjoint GHZ measurements performed at the hub. Let the $\nu$-th state be $n_{\nu}$-GHZ diagonal. Define $u=S-\sum_{\nu=1}^{g}n_{\nu}$ to be the total number of sensors using local probe states. The probe state then can be written as
\begin{align}
    \rho_{\vec{n}}=\left(\bigotimes_{\nu=1}^{g}\sigma_{n_{\nu}}\right)\otimes\ket{+}\bra{+}^{\otimes u},
\end{align}
%If we want, we can order the neighborhoods so that $n_1\geq n_2\dots\geq n_g$. 
where $\vec{n}=(n_1\ n_2\ \cdots\ n_g)$. After interacting with the environment, the state becomes $\rho_{\vec \phi}=U(\vec \phi)\rho_{\vec{n}} U(\vec \phi)^{\dagger}$ where $U(\vec \phi) = e^{-i\vec \phi^{\dagger}\hat{H}}$. The QFI for this snapshot is
\begin{align}
    \label{QFIw/FId}
    \mathcal{F}_\theta(\rho_{\vec{n}})=\frac{(\Delta\lambda)^2 }{S^2}\left(S+\sum_{\nu=1}^{g}\left(C(\vec{x},n_{\nu})n_{\nu}^2-n_{\nu}\right)\right),
\end{align}
where $C(\vec{x},n_{\nu})$ depends on the initial link fidelities, and the size of the GHZ projective measurement. The actual form is 
\begin{align}
        C(\vec{x},n_{\nu})&=\frac{|\bra{\Gamma(n_{\nu})}\sigma_{n_{\nu}}-Z\sigma_{n_{\nu}}Z\ket{\Gamma(n_{\nu})}|^2}{\bra{\Gamma(n_{\nu})}\sigma_{n_{\nu}}+Z\sigma_{n_{\nu}}Z\ket{\Gamma(n_{\nu})}}.
\end{align}
 Here we are letting $\vec{x}$ describe all of Werner parameters of the initial links. If they all have the same Werner parameter $x$, this simplifies to 
\begin{align}
\label{Fcoeff}
C(x,n_{\nu})&=\frac{2^{n_{\nu}}x^{2n_{\nu}}}{(1 - x)^{n_{\nu}} + (1+x)^{n_{\nu}}}.
\end{align}

\subsection{\label{partitioning}GHZ-Projective Measurement Optimization} 

\begin{figure}
\begin{centering}
\subfloat[]{
\includegraphics[width=\linewidth]{ 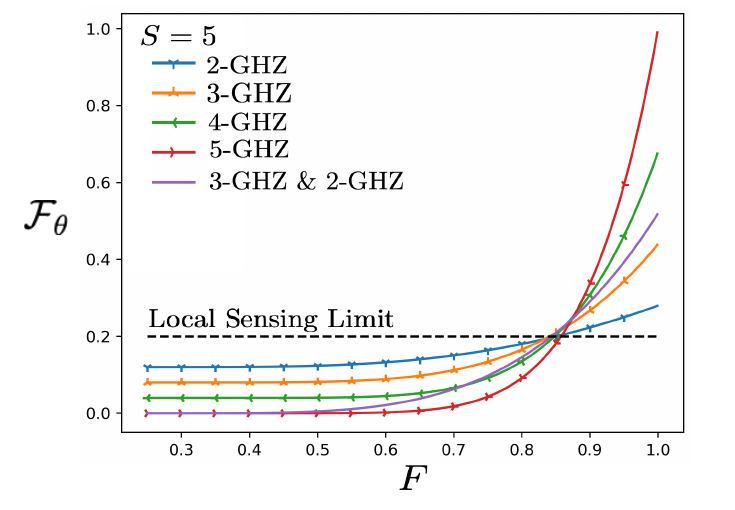}}
\hfill
\subfloat[]{
\includegraphics[width=\linewidth]{ 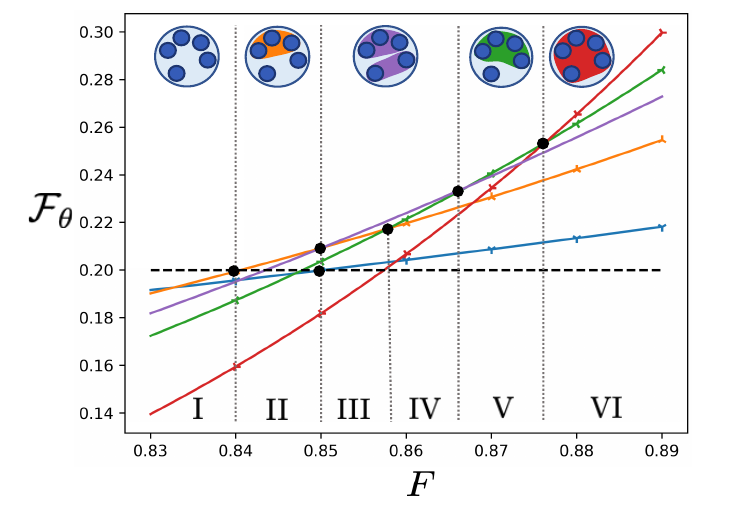}}
\caption{\label{fig:4} (a) QFI vs $F$ for a variety of different initial snapshot probes. Each probe depicted is a possible outcome if there were five successful connections to the hub. A closeup of this plot is shown in (b). The optimal partition of successes is dependent on the value of $F$. The optimal choice of measurements when $m=5$ are represented by the insets at the top of this graph. ($(\Delta\lambda)^2=1$)}
\end{centering}
\end{figure}

\begin{figure}
\includegraphics[width=0.45\textwidth]{ 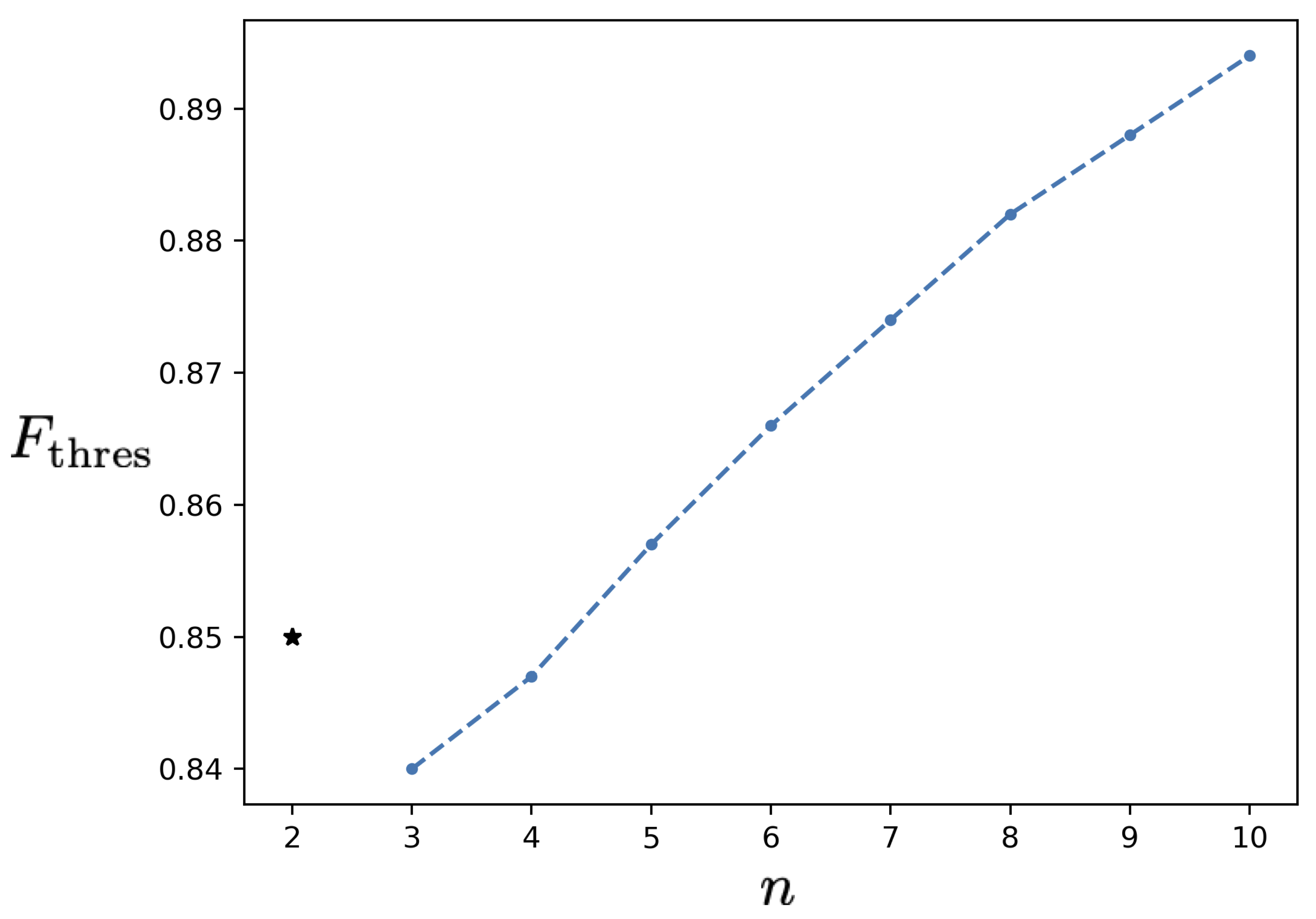}
\caption{\label{fig:5} Threshold fidelity, $F_{\text{thres}}(n)$, below which entanglement doesn't help vs the size of the GHZ projective measurement, $n$. }
\end{figure}

In the previous section it was always optimal to perform the largest possible projective measurement in order to maximize the QFI. Therefore, we always set $n=m$. However, this is not the case when $F<1$. In Fig. \ref{fig:4}, the QFI is plotted against $F$ for five different initial probe states with $S=5$. Assuming that five sensors are able to successfully connect to the hub, these five unique probes states correspond to the five different sets of GHZ projections that the hub can perform. (The trivial case where no GHZ projections are made corresponds to the local sensing limit.) Each GHZ projective measurement has a different threshold fidelity, $F_{\text{thres}}(n)$, and corresponding Werner parameter $x_{\text{thres}}=(4F_{\text{thres}}(n)-1)/3$, below which it is preferable to use local sensors. We can find this Werner parameter for a given $n$ via the expression
\begin{align}\label{thres}
    2^nn(x_{\text{thres}})^{2n}&=(x_{\text{thres}}+1)^n+(1-x_{\text{thres}})^n.
\end{align}
Some values of $F_{\text{thres}}(n)$ are shown in Fig. \ref{fig:5}. For the case that $n=2$, $F_{\text{thres}}(2)>F_{\text{thres}}(3),F_{\text{thres}}(4)$. We conjecture that this is the only exception and in general, $F_{\text{thres}}(n)$ increases with $n$. Note that Eq. \ref{thres} is independent of $S$. Therefore if the system only has two sensors, the sensors benefit from entanglement if $F>0.85$. If the system has more than two sensors, it can benefit from entanglement provided $F>0.84$.
 
Looking at Fig. \ref{fig:4}(b), we see that there is an optimal way to partition the five successful links in order to maximize the QFI for a given $F>0.84$. For all values of $S$, it is optimal to perform the largest GHZ projection possible in the case $F\rightarrow 1$. When $0.84<F<1$, a partition optimization must be solved to determine what the best GHZ projective measurements are at the hub given that there are $m$ successes. 

This problem is parameterized by two variables: $m$, the total number of successfully connected sensors, and $x$, the Werner parameter describing the initial fidelity of these links. We want to partition these $m$ successes into $g$ subsets, $g\in[1,\lfloor\frac{m}{2}\rfloor]$, where the $\nu$-th subset includes $n_{\nu}$ qubits. A GHZ projective measurement is then performed on each of these subsets. The partitioning of $m$ should be chosen to maximize the QFI constrained by  $\sum_{{\nu}=1}^{g}n_{\nu}\leq m$ where $n_{\nu}\in\mathbb{Z}^+\setminus\{1\}$. 

 When $S$ is small, the optimal protocol can be found via an exhaustive search ahead of time. For the case $S=5$, shown in Fig. \ref{fig:4}, the optimal GHZ measurements $\vec{n}$ are defined in Table 1. Using these optimal measurements, Fig. \ref{fig:6}(a) plots average QFI attained with the F-TMBL protocol vs $F$ vs $p$ for a variety of $k$. The optimal measurements for a sampling of larger $m$ and a variety of $F$ values are shown in Table II. Unfortunately, this optimization problem is not convex and therefore can be computationally heavy for large values of $S$.

However, we propose that an exhaustive search can be avoided and that the optimal partitioning will be of the form $\vec{n}=(n_1,\cdots n_{\alpha},n_{\alpha+1},n_g)$  where $n_1=n_2=\cdots=n_{\alpha}=z$ and $n_{\alpha+1}=\cdots=n_g=z-1$ for $z\in[3,m]$. The possible values of $\alpha\in[1,g]$. These values must satisfy $\sum_{\nu=1}^gn_{\nu}=\alpha z+(g-\alpha)(z-1)\leq m$. This heuristic is in agreement with the results of Table II. Generally, when $F\gtrsim0.84$, we find that the optimal partitioning is $\vec{n}=(3,\cdots3)$ where $\alpha=\lfloor m/3\rfloor$. If $m\text{ mod}3=2$, then the optimal partitioning may gain an additional 2-GHZ grouping such that $g\rightarrow g+1$ and $\vec{n}$ becomes $(3,\cdots3,2)$ as the fidelity begins to increase. If the fidelity continues to increase, the optimal grouping of successes will shift so that $\vec{n}=(4,\cdots,3)$ and so forth until $F\sim1$ where the optimal partitioning is $\vec{n}=(m)$. Noting this behavior, we see that $g\in[1,\lceil m/3\rceil]$. 

\begin{table}
\caption{\label{Table1}Optimal $\vec{n}$ for $S=5$. The fidelity regions correspond with those labeled in Fig. ~\ref{fig:4}.}
\begin{tabular}{p{5pt}c|cccccc}
\multicolumn{2}{c}{} & \multicolumn{6}{c}{} \\
     & & $F\in$ I    &  $F\in$ II   & $F\in$ III    & $F\in$ IV    & $F\in$ V     & $F\in$ VI   \\ \cline{2-8}
\multirow{5}{*}{{\ }}
&  $m=2$  & - &  - &  (2) &  (2) &  (2) &  (2) \\
 & $m=3$  & - &  (3) &  (3) &  (3) &  (3) &  (3) \\
  & $m=4$  & - &  (3) &  (3) &  (4) &  (4) &  (4) \\
&   $m=5$ & - &  (3) &  (3, 2) &  (3, 2) &  (4) &  (5) \\
\end{tabular}
\end{table}

\begin{table*}
\caption{\label{Table2}Optimal $\vec{n}$ for $S>5$}
\begin{tabular}{p{5pt}c|ccccccccc}
\multicolumn{2}{c}{} & \multicolumn{9}{c}{} \\
     & & $F\sim0.84$    &  $F=0.86$   & $F=0.88$    & $F=0.9$    & $F=0.92$     & $F=0.94$ & $F=0.96$   & $F=0.98$   & $F=1$      \\ \cline{2-11}
\multirow{5}{*}{{\ }}
&  $m=10$  & (3, 3, 3) &  (4, 3, 3) &  (4, 3, 3) &  (5, 5) &  (5, 5) &  (10)&  (10)&  (10)&  (10) \\
 & $m=15$  & (3, 3, 3, 3, 3) &  (3, 3, 3, 3, 3) &  (4, 4, 4, 3) &  (5, 5, 5) &  (5, 5, 5) &  (8, 7)&  (15)&  (15) &  (15) \\
  & $m=20$  & (3, 3, 3, 3, 3, 3) &  (3, 3, 3, 3, 3, 3, 2) &  (4, 4, 4, 4, 4) &  (5, 5, 5, 5) &  (5, 5, 5, 5) &  (7, 7, 6) & (10, 10)& (20) &(20) \\
%&   $m=25$ & (3, 3, 3, 3, 3, 3, 3, 3) & (4, 3, 3, 3, 3, 3, 3, 3)&  (4, 4, 4, 4, 3, 3, 3) &  (5, 5, 5, 5, 5) &  (7, 6, 6, 6) &  (9, 8, 8) & (13, 12) & (25) & (25) \\
\end{tabular}
\end{table*}

\section{Distillation} 
\begin{figure}
\begin{centering}
\subfloat[Discard Redundant Successes]{
\includegraphics[width=\linewidth]{ 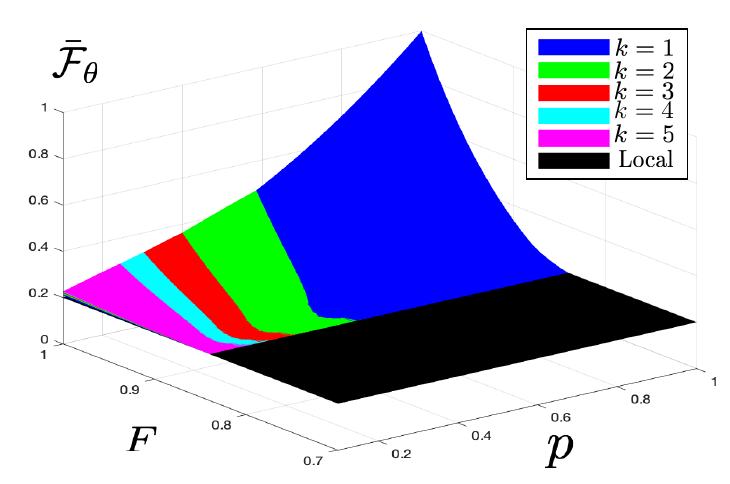}}
\hfill
\subfloat[Distill Redundant Successes]{
\includegraphics[width=\linewidth]{ 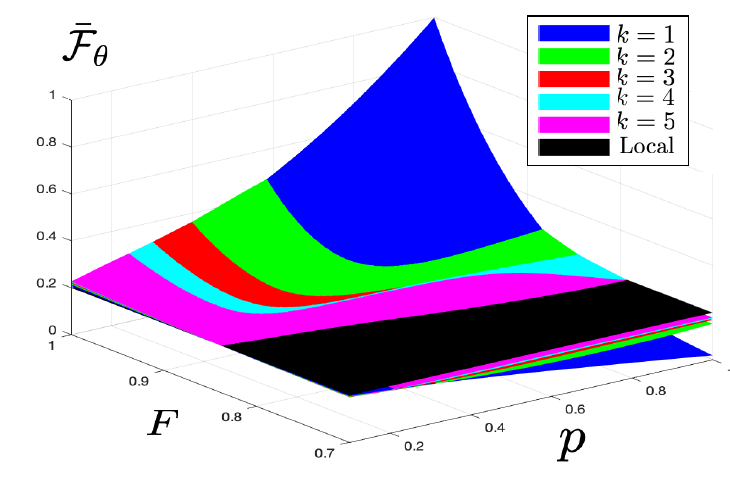}}
\caption{\label{fig:6}  (a) QFI vs $F$ vs $p$ with no distillation and optimal GHZ projections. (b) QFI vs $F$ vs $p$ with with distillation and maximal GHZ projections. Note that the threshold fidelity below which entanglement does not shown an improvement moves to lower $F$ values when distillation is allowed. $S=5$ ($(\Delta\lambda)^2=1$)}
\end{centering}
\end{figure}

Our protocols introduce time multiplexing, so that entanglement generation is attempted some $k$ times before being utilized. In previous sections, the protocols did not make use of multiple successes between sensors and the hub during those $k$ timesteps; they just kept the most recently generated link. In this section, we will show the merit of distilling multiple successful links. 

For this work, each sensor/hub pair that has two or more successful links   will repeatedly use the $2\rightarrow1$ distillation protocol given in \cite{Bennett_1996}. This protocol is probabilistic, but can result in a final state with higher fidelity than the two input states. If the two input states are Werner states with fidelities $F_1$ and $F_2$, the protocol succeeds with probability 
\begin{align}
q(F_1,F_2)=F_1F_2+\frac{F_1(1-F_2)+F_2(1-F_1)}{3}\nonumber\\+\frac{5(1-F_1)(1-F_2)}{9}.
\end{align} 
If successful, we recover a Werner state with fidelity
\begin{align}
F'(F_1,F_2)=\left(F_1F_2+\frac{(1-F_1)(1-F_2)}{9}\right)q^{-1}.
\end{align} 

If a sensor and hub pair has $l$ successful links where $l>1$, they will repeatedly perform the distillation explained above on states with the same fidelity. If a round of distillation fails, the sensor/hub pair might still have a lower fidelity entangled link that it can use for sensing. However, the protocol will always try and use the higher fidelity state. If no distillation attempts succeed and $l$ was even, the sensor must sense locally. However, if no distillation attempts succeed and $l$ was odd we can choose to \textit{keep} the remaining link with fidelity $F$ or to \textit{discard} it. For the main part of this work, we choose to discard it because it is preferably when $F$ is near $F_{\text{thres}}(3)$. A pseudocode is shown in Algorithm \ref{EPSA} to calculate the possible fidelities and their probability of occurring for all outcomes of this protocol. In Appendix \ref{Appendix:Dist} we show results for the other option where we keep the the remaining link.

We do not claim this to be the best distillation technique, but employ this simple scheme in order to prove that distillation can lower the threshold fidelity $F_{\text{thres}}(3)$ over which entanglement shows an improvement over local sensors as shown in Fig. \ref{fig:6}. For this figure, the maximal GHZ projective measurement is always performed at the hub. The optimal GHZ projective measurements are even more difficult to find now because different sensor/hub pairs can now have different fidelities resulting from distillation. Optimizing the size of these GHZ projections is an open area of interest. This protocol with distillation always outperforms that without distillation when $F<1$.
%When distillation is allowed for, the fidelity of the entangled links can vary.The fidelity of the Bell pairs shared by the hub and the $s$-th sensor just prior to the hub's projective measurements will be denoted to as $F_s$ where  $1\leq s\leq S$. We can write $\vec{F}=(F_1,F_2,...,F_S)$. Otherwise, it can be assumed that $F$ refers to a common fidelity that all the Bell states initially are prepared at.
\begin{algorithm}[H]
  \caption{Nested $2\rightarrow1$ Distillation}
  \label{EPSA}
   \begin{algorithmic}[1]
   \State \textbf{def} DistLink$(F,l,prb):$ 
       \If{$l=0$}\Comment{No links remain endcase}
       \State{\textbf{yield}$[0,prb]$} \EndIf
       \If{$l=1$ \textbf{ or } $F\leq0.5$}\Comment{Can't distill endcase}
       \State{\textbf{yield}$[F,prb]$} 
       \Else
       \State{$SucPrb=F_1F_2+\frac{F_1(1-F_2)+F_2(1-F_1)}{3}+\frac{5(1-F_1)(1-F_2)}{9}$} \Comment{Success prob of distillation}
       \State{$SucF=\left(F_1F_2+\frac{(1-F_1)(1-F_2)}{9}\right)/SucPrb$} \Comment{New fidelity}
        \State{$MaxSuc=\lfloor l/2\rfloor$} 
        \For{$s$ in \textbf{range}$(0,MaxSuc)$} \Comment{Iterates over outcomes}
        \State{$t={MaxSuc\choose s}prb(SucPrb)^{s}(1-SucPrb)^{MaxSuc-s}$}
        \State{\textbf{yield from} DistLink($SucF,s,t)$}
        \EndFor
       \EndIf
    \State{}
    \State{DistLink$(F,l,1)$} \Comment{Function initialization}
   \end{algorithmic}
\end{algorithm}

\section{Measurement Designs}
\begin{figure}
\begin{centering}
\subfloat[]{
\includegraphics[width=\linewidth]{ 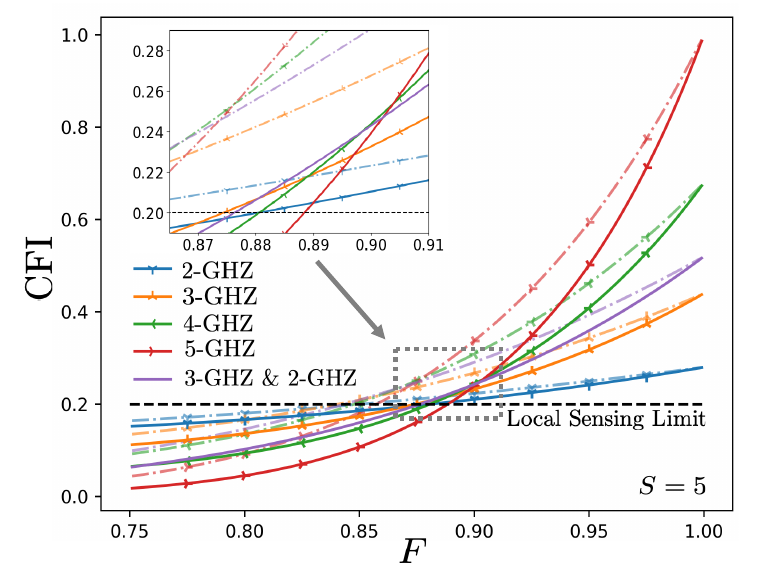}}
\hfill
\subfloat[]{
\includegraphics[width=\linewidth]{ 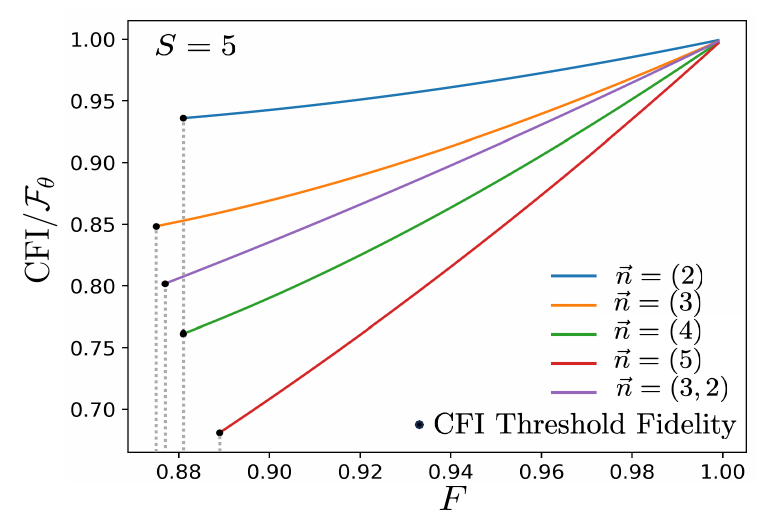}}
\caption{\label{fig:7} (a) Plot of the CFI vs $F$ for a variety of probe states and measurement schemes. The semi-transparent dashed lines are for the QFI attaining global measurement while the solid lines correspond to maximum results of the sub-optimal local measurement scheme. (The CFI of these measurements are dependent on the values of the parameters, and may be as low as zero.) The inset is a closeup of this plot, highlighting the behavior where the CFI of the local measurement scheme intersects the local sensing limit. (b) Plot of maximum CFI/QFI v $F$ for the local measurement scheme the black dots correspond to the CFI threshold fidelity, below which entangled probe states with local measurements perform worse than product states. ($(\Delta\lambda)^2=1$)}
\end{centering}
\end{figure}
Up until this point, this paper has been focused on comparing the quantum Fisher information associated with different probe states, without accounting for how one actually measures the state after it interacts with the environment. One way to determine an optimal measurement scheme to calculate the symmetric logarithmic derivative (SLD) operator, as we have done in Appendix \ref{Appendix:SLD}. The eigenbasis for this operator is a QFI attaining measurement, although it is not necessarily the only one. Let $n_{\nu}$ be equal to the size of the $\nu$-th GHZ projective measurement and $n_{\nu}=1$ for all $\nu\in[g+1,g+u]$. The set of POVMs are 
\begin{align}
\Pi_{\nu,d}=U(\vec{\phi})\bigotimes_{\nu=1,d\in\{0,1\}}^{g+u}\ket{c(n_{\nu},d)}\bra{c(n_{\nu},d)}U(\vec{\phi})^{\dagger},
\end{align}
where
\begin{align}
\ket{c(n_{\nu},d)}=\frac{1+(-1)^d i}{2}\ket{0}^{\otimes n_{\nu}}+\frac{1-(-1)^di}{2}\ket{1}^{\otimes n_{\nu}},
\end{align}
and $\Pi_0=\mathbb{I}-\sum{\Pi_{\nu,d}}$. These optimal measurements, which are used to estimate the the unknown parameter, are also dependent on $\vec{\phi}$ (and therefore $\theta$). This is not uncommon in estimation theory problems and can be worked around by sequential sensing method such as those proposed in \cite{hayashi_introduction_2015}. Unfortunately, we need access to additional entanglement, a limited resource we are trying to use efficiently in our sensing protocol, to perform the joint measurements proposed above. This is consistent with the optimal measurement protocol in \cite{PhysRevA.97.042337} which  made use of GHZ states to perform measurements.

Instead, we can consider performing local measurements on all of our sensing qubits. Lets assume that all the intial links are of the same initial fidelity and therefore have the same Werner parameter $x$. We can order the qubits such that the  $n_1$ qubits in the $n_1$-GHZ diagonal state are at sensors $\{1,\cdots,n_1\}$ and the  $n_2$ qubits in the $n_2$-GHZ diagonal state are at sensors $\{n_1+1,\cdots,n_1+n_2\}$  and so forth. To more easily refer to qubits in a certain GHZ-diagonal state, lets define 
\begin{align}
    Q_{\nu}=\sum_{i=1}^{\nu-1}n_i,
\end{align}
and define $Q_1=0$. The qubits at sensors $\{S-u,\cdots,S\}$ are all local probes where $u=S-\sum_{\nu=1}^gn_{\nu}$. The classical Fisher information (CFI) associated with measuring the probe states in the $\ket{+},\ket{-}$ basis is,
\begin{align}
    \frac{(\Delta\lambda)^2}{S^2}\left(u+\sum_{\nu=1}^g\frac{x^{2S}\sin^2(\Delta\lambda\sum_{q=1+Q_{\nu}}^{n_{\nu}+Q_{\nu}}\phi_q)}{1-x^{2S}\cos^2(\Delta\lambda \sum_{q=1+Q_{\nu}}^{n_{\nu}+Q_{\nu}}\phi_q)}n_{\nu}^2\right).
\end{align}

 When $x=1$, so all the Werner states are actually pure Bell states, the dependence on the unknown parameters can be canceled out, and our expression for CFI reduces to the QFI expression found in Eq. \ref{unit-fidelity qfi}. This means that when the Bell pairs connecting the sensors to the hub are pure, a local measurement scheme can achieve the QFI. In general the CFI depends on the values of the unknown parameters. The CFI is largest when $\sum_{q=1+Q_{\nu}}^{n_{\nu}+Q_{\nu}}\phi_q=\frac{(z_{\nu}+1/2)\pi}{\Delta\lambda}$ for all $\nu\in[1,g]$ where $z_{\nu}\in\mathbb{Z}$. When $F<1$, the CFI associated with the local measurement scheme is less than the QFI even when the measurement is optimized for the parameters, as shown in Fig. \ref{fig:7}. There is a range of $F$ where performing local measurements on entangled states outperforms using local probes, however the required initial $F$ for the Werner states is higher than the threshold $F$ required by the QFI calculations.  The optimal partitioning of successes follows a similar behavior to that shown in Fig. \ref{fig:4}. The derivation for the above expression can be found in Appendix \ref{Appendix:LocalMeasurement} along with a more general expression for when the Werner states have different fidelities.

\section{Conclusion}
In this paper we modeled a star shaped quantum network focused on estimating the average phase parameter over all the sensors. Sensors were able to probabilistically generate entangled Werner states with the central hub, after which more complicated GHZ states could be prepared to use as probe states. Three different sensing protocols were introduced in this paper: \textit{Immediate Sensing}, \textit{Fixed Time Multiplexing Block Length} (F-TMBL), and \textit{Variable Time Multiplexing Block Length} (V-TMBL). The average quantum Fisher information over time was calculated in order to compare which of these protocols best utilized the available entanglement as a resource for different network conditions. When the links have unit fidelity, the QFI is maximized when the largest GHZ state possible is created. If the link success probability $p<2-\sqrt{2}$, time multiplexing helps because it allows successful links from different timesteps to be combined into GHZ states to be used in sensing. For the case F-TMBL protocol, an optimal $k$ can be chosen that is independent of the number of sensors. For the V-TMBL protocol, an optimal $\mu$ can be chosen, which does change with the number of sensors. As expected, the V-TMBL does as well or better than the F-TMBL protocol as it has more flexibility, although the improvement diminishes as the number of sensors is increased. 

When the initial fidelity is less than one, it is not always preferable to prepare one large GHZ state from any of the successful links. In fact, the optimal partitioning of $m$ successes is dependent on the fidelity value. If $F<0.84$, entangled probe states have a lower QFI value than local probe states. For $0.84<F<1$, we showed the optimal protocol for a five sensor network, and discussed a heuristic for finding the optimal partitioning for larger networks. However, we do not provide a proof of optimality and further work may be done to develop an algorithm for determining the optimal partitioning. We show that distillation schemes do help improve the region of fidelities over which our F-TMBL protocol outperforms local sensing by evaluating a simple nested scheme. Further work could be done on combining the optimal partitioning of successes with distillation and other distillation schemes could be investigated. Additionally, evaluating the performance of the V-TMBL protocol with such inclusions may be fruitful.

We found a QFI attaining measurement strategy. Unfortunately, this set of POVMs depend on the values of the parameters themselves and require use of entanglement to perform global measurements. Since entanglement is the limiting resource, we also propose a non-optimal local measurement strategy that outperforms using non-entangled states for a range of fidelities. Finding better measurements or using entanglement as a resource for performing measurements are other areas that could be investigated in future work. 

Our current model can be expanded to include memory decoherence and also can be evaluated over a wider range of network topologies. Additionally, the protocols presented here can be applied to more specific sensing problems. Although we do not prove optimality on our proposed protocols, we believe that we highlight many interesting areas of consideration and demonstrate the importance of using quantum Fisher information as a metric for quantum networks.

\section{Acknowledgements}
This work was co-funded by AFOSR grant number FA9550-22-1-0180, and the National Science Foundation (NSF) grant
CNS-1955834 and NSF ERC Center for Quantum Networks (CQN) grant EEC-1941583.

\bibliography{main}
\appendix
\onecolumngrid

\section{\label{Appendix:Latency}Latency and Memory Requirements}
\begin{figure}
\includegraphics[width=\textwidth]{ 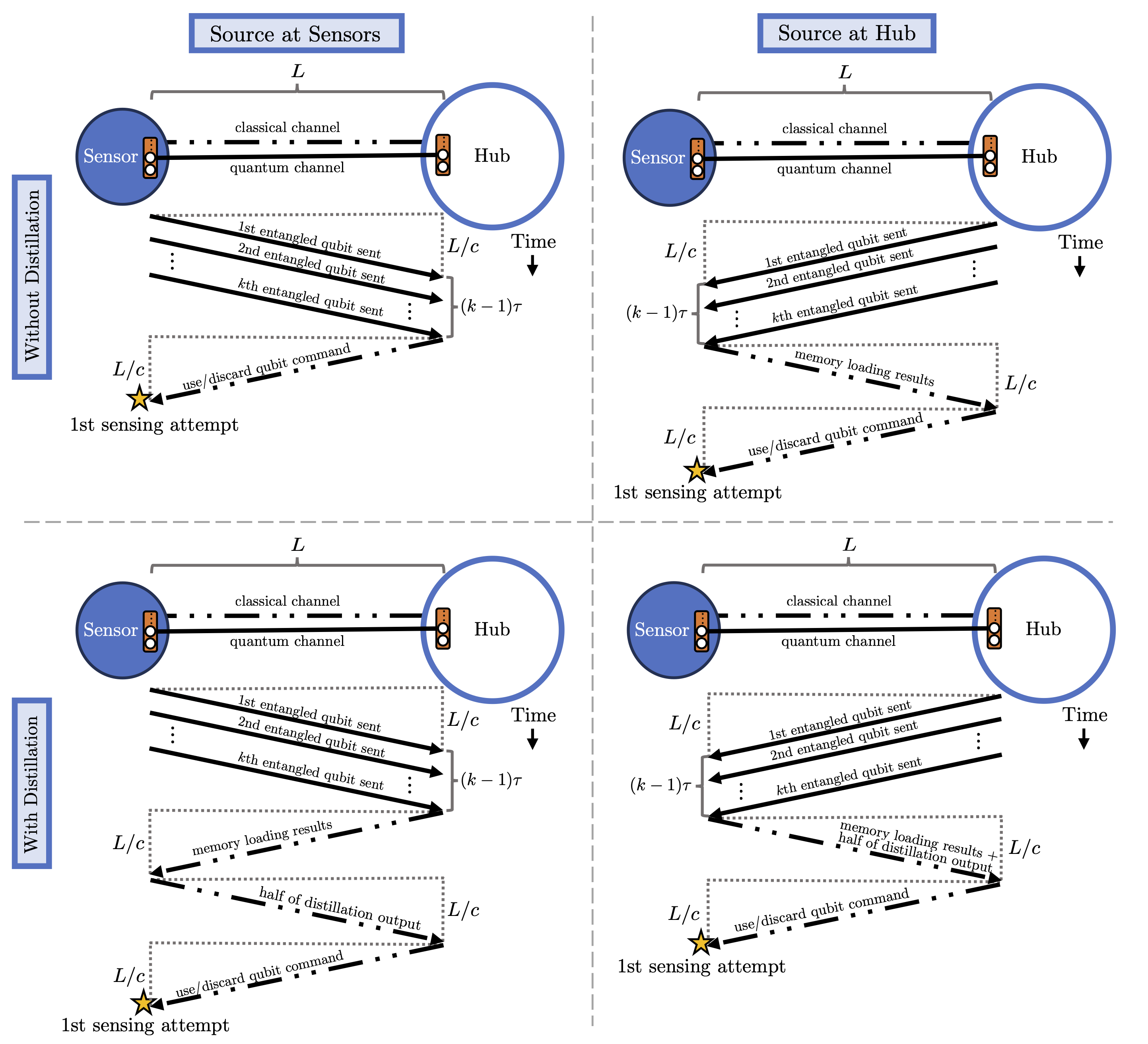}
\caption{\label{fig:8}Timing diagrams. Without distillation, the latency is lower when the entanglement source is at the sensors. With distillation, the latency is lower when the entanglement source is at the hub. If the sensors are all a distance $L$ away from the hub, the latencies are $2L/c+(k-1)\tau$ and $3L/c+(k-1)\tau$ respectively.}
\end{figure}
%Timing diagram for our protocol with and without distillation. Without distillation, the latency is lower when entanglement originate in the sensors. With distillation, the latency is lower when entanglement originate at the hub. If the sensors are all a distance $L$ away from the hub, the latencies are $2L/c+(k-1)\tau$ and $3L/c+(k-1)\tau$ respectively.
The  latency and number of quantum memories required at the sensors and at the central hub are dependent on the physical implementation. In this section, we will evaluate two simple models where entanglement is either initially generated at the sensors or at the hub. The results of these two models are dependent on whether the protocol requires entanglement distillation. For all the cases explored below, lets assume that all of the sensors are some distance $L$ away from the central hub as shown in Fig.~\ref{fig:7}. We will denote $\tau$ as the reset time needed between entanglement generation attempts at the sensor/hub.
\subsection{No Distillation}
\subsubsection{Case 1: Initial Generation at Sensors }
First, lets assume the initial entangled pairs are generated at the sensor nodes. This may be done with a color center memory \cite{ruf_quantum_2021} so that, one qubit will be generated locally at the sensor initially via electron spin then stored more securely as a nuclear spin, while the other will be emitted in the form of a photo along an optical channel connecting the sensor and the hub.  It will take $L/c$ seconds for the photon to arrive at the hub, where another Bell pair will have been created in a similar fashion. A BSM can be performed on the two photonic qubits in such a way  that the remaining two qubits, one held at the sensor and one at the hub, are entangled with some probability $p$. The success/failure of this measurement can be heralded by the hub and communicated back to the sensor in real time over a classical channel. If the protocol in question has additional time multiplexing, the hub will know about the $k$-th success/failure at time $L/c+(k-1)\tau$. At this point the hub  can decide to perform a GHZ projection on a subset of the successfully stored qubits. We will assume that the projection occurs instantaneously compared to the other time scales.  It then communicates back to the sensors which qubit they should use to sense with or if they need to prepare local probes states.  
The sensors finally can use their stored qubits to sense with after a latencies of $2L/c+(k-1)\tau$.  For this scenario, each sensor must have at least $\lceil\frac{2L}{c\tau}\rceil+k$ quantum memories and the hub must have $Sk$ quantum memories.
\subsubsection{Case 2: Initial Generation at Hub }
If instead the original entangled pair originates at the hub, the latency increases. This is because after the qubits are loaded onto the memories at the sensors, the sensors must communicate to the hub which succeeded. Then, the hub can perform its GHZ projective measurements and communicate back to the sensors whether they should use their entangled qubit or prepare a local probe. In this scenario, the latency becomes $3L/c+(k-1)\tau$ and each sensor must have at least $\lceil\frac{2L}{c\tau}\rceil+k$ quantum memories and the hub must have $S(\lceil\frac{2L}{c\tau}\rceil+k)$ quantum memories.

\subsection{With Distillation}  
\subsubsection{Case 1: Initial Generation at Sensors }
Following the steps explained in the previous section, if the initial entangled pair is generated at the sensors, it is not until $L/c+(k-1)\tau$ seconds after beginning that the hub has attempted to store $k$-th qubit. At this point, if it has successfully stored one or more entangled qubits from more than two sensors, the hub can perform half of a distillation circuit on the qubits it holds locally and communicate what it measures to the respective sensors. The sensors will know which qubits the hub successfully stored after $2L/c+(k-1)\tau$ seconds and can run its half of the distillation circuit before comparing measurements. At this point the sensor must communicate back to the hub whether the distillation circuit succeeded and it is not until  $3L/c+(k-1)\tau$ seconds that the hub can perform a GHZ projective measurement. The sensors must wait another $L/c$ to find out if they can use the initial qubit to sense with. Under these conditions, there is a latency of $4L/c+(k-1)\tau$ and each sensor must have at least $\lceil\frac{4L}{c\tau}\rceil+k$ quantum memories, and the hub must have $S(\lceil\frac{2L}{c\tau}\rceil+k)$ quantum memories.
\subsubsection{Case 2: Initial Generation at Hub }
Now lets assume the initial entangled pairs are generated at the hub nodes. After the initial entangled pair is generated, it takes $L/c+(k-1)\tau$ seconds for the sensors to attempt to store the $k$ qubits. The sensors then can perform half of a distillation circuit on the qubits they successfully stored, and communicate what they measure to the respective hub. The hub will know which qubits the have been successfully stored after $2L/c+(k-1)\tau$ seconds and can run its half of the distillation circuit before comparing measurements. At this point the hub can perform a GHZ projective measurement and communicate to the sensors whether to use their entangled qubits or to prepare local states to sense with. Under these conditions, there is a latency of $3L/c+(k-1)\tau$ and each sensor must have at least $\lceil\frac{2L}{c\tau}\rceil+k$ quantum memories and the hub must have $S(\lceil\frac{2L}{c\tau}\rceil+k)$ quantum memories. 

\section{\label{Appendix:GHZcoeff}GHZ coefficients} 
Let there be $n$ Werner states, where the $s$-th state can be written $\beta_{A_sB_s} = x_s \Phi_{A_sB_s} + \frac{(1-x_s)}{4}\mathbb{I}_{A_s}\otimes\mathbb{I}_{B_s}$ where $\Phi_{A_sB_s}$ corresponds to a Bell pair between two qubits $A_s$ and $B_s$. The Werner parameter of this state is $x_s=\frac{4F_s-1}{3}$ where $F_s$ is the fidelity. Afterwards, one qubit of each pair will be measured off to prepare the state:
\begin{equation}
 \sigma_{n}=\bra{\Gamma(n)}_{B_1\cdots B_{n}}\bigotimes_{s=1}^{n}\beta_{A_sB_s}\ket{\Gamma(n)}_{B_1\cdots B_{n}}\nonumber.
\end{equation}

The coefficients making up the different components of $\rho_n$ were calculated in \cite{kaur2023entanglement} \cite{10313684}. The $n$-GHZ basis vectors can be written as $\ket{\psi_{j,i_1,i_2\cdots i_{n-1}}}$, where
\begin{equation}
\label{ghz-basis}
    \ket{\psi_{j,i_1,i_2\cdots i_{n-1}}} = \bigotimes_{\alpha=1}^{n-1}X_\alpha^{i_\alpha}Z_1^j\ket{\Gamma(n)},
\end{equation}
where $i_{\alpha},j\in [0,1]$ and $\ket{\Gamma(n)}$ was previously defined in Eq. \ref{eq:GHZdef}. 
In this notation the basis for a $3$-qubit system is given as:
\begin{align*}
    \ket{\psi_{0,0,0}} &= \ket{\Gamma(3)}\quad\quad\quad\  \ket{\psi_{1,0,0}} = Z_1\ket{\Gamma(3)}\\\ket{\psi_{0,1,0}} &= X_1\ket{\Gamma(3)}\quad\quad \ket{\psi_{1,1,0}} = X_1Z_1\ket{\Gamma(3)}\\\ket{\psi_{0,0,1}} &= X_2\ket{\Gamma(3)}\quad\quad \ket{\psi_{1,0,1}} = X_2Z_1\ket{\Gamma(3)}\\
    \ket{\psi_{0,1,1}} &= X_1X_2\ket{\Gamma(3)}= X_3\ket{\Gamma(3)}\\ \ket{\psi_{1,1,1}} &= X_1X_2Z_1\ket{\Gamma(3)} = X_3Z_1\ket{\Gamma(3)}.
\end{align*}
This notation can be further abbreviated as $\ket{\psi_{j\vec{i}}}$ where $\vec{i}=(i_1,i_2,\cdots i_{n-1})^{T}$.

The coefficients for $\ket{\psi_{0,\vec{0}}}\bra{\psi_{0,\vec{0}}}$: 
\begin{multline}\label{DiffFGHZ}
    E_{0,\vec{0}}=\prod_{s=1}^n x_s + \sum_{s=1}^{n-2} \frac{1}{2^{s+1}}\frac{1}{s!}\sum_{a_1\in\left[1,n\right]}\sum_{a_2\in\left[1,n\right]\backslash a_1}\cdots\sum_{a_s\in\left[1,n\right]\backslash\prod_{b= 1}^{s-1}a_b}
    \Pi_{c=a_1}^{a_{s}}(1-x_c) \ \Pi_{b=[1,n]\backslash\Pi_{b=1}^{s}a_b}x_b\\+ \frac{1}{2^n}\sum_{s=1}^nx_s\Pi_{a\in[1,n]\backslash s}(1-x_a)+\frac{1}{2^n}\Pi_{s=1}^n (1-x_s)^n .
\end{multline}

The coefficients for $\ket{\psi_{1,\vec{0}}}\bra{\psi_{1,\vec{0}}}$ are: 
\begin{multline}
 E_{1,\vec{0}}=\sum_{s=1}^{n-2} \frac{1}{2^{s+1}}\frac{1}{s!}\sum_{a_1\in\left[1,n\right]}\sum_{a_2\in\left[1,n\right]\backslash a_1}\cdots\sum_{a_s\in\left[1,n\right]\backslash\Pi_{b= 1}^{s-1}a_b}\Pi_{c=a_1}^{a_{s}}(1-x_c) \ \Pi_{b=[1,n]\backslash\Pi_{b=1}^{s}a_b}x_b \\+\frac{1}{2^n}\sum_{s=1}^nx_s\Pi_{a\in[1,n]\backslash s}(1-x_a)+\frac{1}{2^n}\Pi_{s=1}^n (1-x_s)^n \label{eq:GHZ-coeffi-different}
\end{multline}

The coefficients for $X_{\alpha_y}\cdots X_{\alpha_2}X_{\alpha_1}\ket{\psi_{0,\vec{0}}}\bra{\psi_{0,\vec{0}}}X_{\alpha_1}X_{\alpha_2}\cdots X_{\alpha_y}$ 

and $X_{\alpha_y}\cdots X_{\alpha_2}X_{\alpha_1}\ket{\psi_{1,\vec{0}}}\bra{\psi_{1,\vec{0}}}X_{\alpha_1}X_{\alpha_2}\cdots X_{\alpha_y}$ are: 
\begin{multline}
 E_{0/1,\vec{i}}=\sum_{s=y}^{n-2} \frac{1}{2^{s+1}}\frac{1}{(s-y)!}\sum_{a_1\in\left[1,n\right]}\sum_{a_2\in\left[1,n\right]\backslash a_1}\cdots\sum_{a_i\in\left[1,n\right]\backslash\Pi_{b= 1}^{s-1}a_b}\Pi_{s=1}^b\delta_{a_s = \alpha_s} \ \Pi_{c=a_1}^{a_{i}}(1-x_c) \ \Pi_{b=[1,n]\backslash\Pi_{b=1}^{s}a_b}x_b\\+\frac{1}{2^n}\sum_{s=1}^nx_s\Pi_{a\in[1,n]\backslash s}(1-x_a) +\frac{1}{2^n}\Pi_{s=1}^n (1-x_s)^n. \label{eq:GHZ-X-coeffi-different}
\end{multline}

For the above equations, $[1,n]$ denotes the set of integer values spanning from 1 to $n$. It is of note that the coefficients for states that vary only by a $Z$ gate are the same in all cases except for $\op{\psi_{0,\vec{0}}}$ and $\op{\psi_{1,\vec{0}}}$, i.e. $E_{0,\vec{i}}=E_{1,\vec{i}} \ \forall \vec{i}\neq\vec{0}$. When all the fidelities approach one, we see that these expressions simplify properly so that the coefficient corresponding to $\op{\psi_{0,\vec{0}}}$ goes to one while the rest go to zero. 

If $x_1=x_2=\cdots=x_S=x$ we can rewrite the summations as
\begin{align}
E_{0,\vec{0}}&=\frac{1}{2}\left(x^S+\frac{1}{2^S}\left((1+x)^S+(1-x)^S\right)\right)\\
E_{1,\vec{0}}&=\frac{1}{2}\left(-x^S+\frac{1}{2^S}\left((1+x)^S+(1-x)^S\right)\right),
\end{align}
which agree with results found in \cite{ma_quantum_2011}.

\section{\label{QFI_Calc}QFI Calculation}
The single parameter quantum Cramer-Rao bound Eq. \ref{QCRB-single}  has a multi-parameter counterpart. If we are trying to estimate all the values of $\phi_s$ for $s
\in\{1\cdots S\}$ with a probe state $\rho$ and have unbiased estimators, the covariance matrix $\text{Cov}(\vec{\phi},\rho)$ satisfies the following inequality:
\begin{align}
    \text{Cov}(\vec{\phi},\rho)\geq\mathcal{F}_{\vec{\phi}}^{-1}(\rho),
\end{align}
where $\mathcal{F}_{\vec{\phi}}(\rho)$ now is a $S$ by $S$ matrix called the quantum Fisher information matrix (QFIM) consisting of elements given by
\begin{align}
\label{1parameterQFI}[\mathcal{F}_{\vec{\phi}}(\rho)]_{ab}=4\sum_{\gamma',\gamma}E_{\gamma}\left(\frac{E_{\gamma'}-E_{\gamma}}{E_{\gamma'}+E_{\gamma}}\right)^2
\left\langle\psi_{\gamma'}U(\vec \phi)^{\dagger} \middle|{\partial_{\phi_a}U(\vec \phi)\psi_{\gamma}}\right\rangle\left\langle\partial_{\phi_b}\psi_{\gamma}U(\vec \phi)^{\dagger} \middle|U(\vec \phi)\psi_{\gamma'}\right\rangle,
\end{align}
where $\rho=\sum_{\gamma}E_{\gamma}\ket{\psi_{\gamma}}\bra{\psi_{\gamma}}$ is the probe state expressed in its eigenbasis and $\partial_{\phi_a}=\frac{\partial}{\partial\phi_a}$. First, we will calculate the QFIM of $\vec{\phi}$. To do so we will make use of the QFIM property \cite{liu_quantum_2020} that
\begin{align}\label{cross}
    \mathcal{F}(\rho)=\sum_{i}\mathcal{F}(\rho_i) \ \ \ \text{if } \rho=\bigotimes_i\rho_i,
\end{align}
We can then transform this QFIM in order to get the QFI when estimating $\theta$ .When the parameter of interest $\theta$ is a function of $\vec{\phi}$, then the QFIMs with respect to  $\theta$ is
\begin{align}
    \mathcal{F}_{\theta}(\rho)= \hat{J}\mathcal{F}_{\vec{\phi}}(\rho)\hat{J}^{\dagger},
\end{align}
where $\hat{J}$ is the Jacobian operator, defined as $J_{i} = \partial \theta/\partial \phi_i$ \cite{liu_quantum_2020}. 

\subsection{QFIM of $\vec{\phi}$} The probe state in our paper may consist of $g$ different GHZ diagonal states -- corresponding to the $g-$disjoint GHZ measurements performed at the hub. Let the $\nu$th GHZ state be $n_{\nu}$-GHZ diagonal. Lets define $u=S-\sum_{\nu=1}^{g}n_{\nu}$ to be the total number of sensors using local probe states. The probe state then can be written as:
\begin{align}
    \rho_{\vec{n}}=\left(\bigotimes_{\nu=1}^{g}\sigma_{n_{\nu}}\right)\otimes\ket{+}\bra{+}^{\otimes u},
\end{align}
where $\vec{n}=(n_1\ n_2\ \cdots\ n_g)$ and $\sigma_{n_{\nu}}$ is a GHZ diagonal state whose explicit form can be found in Appendix \ref{Appendix:GHZcoeff}.  For our problem, the indexing of the parameters/sensors is arbitrary so we can order the qubits such that the  $n_1$ qubits in the $n_1$-GHZ diagonal state are at sensors $\{1,\cdots,n_1\}$ and the  $n_2$ qubits in the $n_2$-GHZ diagonal state are at sensors $\{n_1+1,\cdots,n_1+n_2\}$  and so forth. To more easily refer to qubits in a certain GHZ-diagonal state, let 
\begin{align}
    Q_{\nu}=\sum_{i=1}^{\nu}n_i,
\end{align}
where we will define $Q_0=0$. With this notation, qubits at sensors $\{S-u,\cdots,S\}$ are all local probes.  After interacting with the environment, the state becomes $\rho_{\vec \phi}=U(\vec \phi)\rho_{\vec{n}} U(\vec \phi)^{\dagger}$ where $U(\vec \phi) = e^{-i\vec \phi^{\dagger}\hat{H}}$.  

Since the form of the probe state is a tensor product of states, we will utilize  Eq. \ref{cross} to get
\begin{align}
    [\mathcal{F}_{\vec{\phi}}
    (\rho_{\vec{\phi}})]_{ab}=\sum_{\nu=1}^g\mathcal{F}_{\vec{\phi}}(V_{\nu}\sigma_{n_\nu}V_{\nu}^{\dagger})+\sum_{\nu=S-u}^S\mathcal{F}_{\vec{\phi}}(e^{-i\hat{h}_{\nu}\phi_{\nu}}\ket{+}_{\nu\ \nu}\bra{+}e^{i\hat{h}_{\nu}\phi_{\nu}}),
\end{align}
where
\begin{align}
    V_{\nu}=\prod_{q=1+Q_{\nu-1}}^{Q_{\nu}}e^{-i\hat{h}_q\phi_q}\bigotimes_{q'=1+Q_{\nu-1},\  q'\neq q}^{Q_{\nu}}\mathbb{I}_{q'},
\end{align}
is just the action of the unitary of $U(\vec{\phi})$ acting on the subsystem of qubits in the $\nu$-th GHZ diagonal state.  

Focusing first on the contribution of the entangled states, we will solve for $[\mathcal{F}(V_{\nu}\sigma_{n_\nu}V_{\nu}^{\dagger})]_{ab}$. Looking at Eq. \ref{1parameterQFI}, we see that  
\begin{align}
\left\langle\psi_{j'\ \vec{i'}}V_{\nu}^{\dagger}\middle|\partial_{\phi_a}V_{\nu}\psi_{j\ \vec{i}}\right\rangle=
\begin{cases}
       -i\bra{\psi_{j'\ \vec{i'}}}\hat{h}_{a}\ket{\psi_{j\ \vec{i}}} &\quad\text{if } a \in \{1+Q_{\nu-1},\dots, Q_{\nu}\} \\
       \text{0} &\quad\text{otherwise.} \\ 
\end{cases}
\end{align}

We can write $\hat{h}_a$ in terms of $Z_a$, where $Z_a$ applies a $Z$-gate to the $a$th qubit and applies an identity to all others. Then $\hat{h}_a=-\frac{\Delta\lambda}{2}\hat{Z}_a+(\lambda_{0}+\frac{\Delta\lambda}{2})\hat{\mathbb{I}}$, which satisfies $\hat{h}_a\ket{0}=\lambda_{0}\ket{0}_a$ and $\hat{h}_a\ket{1}=\lambda_{1}\ket{1}_a$ where $\Delta\lambda=\lambda_{1}-\lambda_{0}$. The first term is only nonzero when $\ket{\psi_{j'\ \vec{i}'}}=\hat{Z}_a\ket{\psi_{j\ \vec{i}}}$ since 
\begin{align}
-\frac{\Delta\lambda}{2}\bra{\psi_{j'\ \vec{i}'}}\hat{Z}_a\ket{\psi_{j\ \vec{i}}}=-\frac{\Delta\lambda}{2}\bra{\psi_{j\ \vec{i}}}\hat{Z}_a\hat{Z}_a\ket{\psi_{j\ \vec{i}}}\nonumber=-\frac{\Delta\lambda}{2}\bra{\psi_{j\ \vec{i}}}\hat{\mathbb{I}}\ket{\psi_{j\ \vec{i}}}=-\frac{\Delta\lambda}{2}.
\end{align}
Therefore this term only contributes if  $j\neq j'$ and $\vec{i}=\vec{i}'$. For all $a \in \{1+Q_{\nu-1},\dots, Q_{\nu}\}$, $Z_a\ket{\psi_{j\ \vec{i}}}$ is the same state, as it doesn't matter which qubit the $Z$ gate is applied to. 

The second term, $(\lambda_{0}+\frac{\Delta\lambda}{2})\bra{\psi_{j'\ \vec{i}'}}\psi_{j\ \vec{i}}\rangle$, is only
 nonzero iff $j=j'$ and $\vec{i}=\vec{i}'$. However, if this is the case then $E_{j'\ \vec{i}'}-E_{j\ \vec{i}}=E_{j\ \vec{i}}-E_{j\ \vec{i}}=0$, so the coefficient in the front of Eq. \ref{1parameterQFI} will be zero. Thus, this second term does not contribute to the calculation. 

Now we note the form of $\sigma_{n_\nu}$ defined in \ref{Appendix:GHZcoeff}, $E_{j'\ \vec{i}}=E_{j\ \vec{i}}$  unless $\vec{i}=\vec{0}$. So if   $\vec{i}\neq\vec{0}$, then $E_{j'\ \vec{i}}-E_{j\ \vec{i}}=0$. From this we find the QFIM to be
\[   
[\mathcal{F}_{\vec{\phi}}(V_{\nu}\sigma_{n_\nu})V_{\nu}^{\dagger}]_{ab} = 
     \begin{cases}
       C(\vec{x},n_{\nu})(\Delta\lambda)^2 &\quad\text{if  } a,\ b\in \{1+Q_{\nu-1},\dots, Q_{\nu}\} \\
       \text{0} &\quad\text{otherwise,} \\ 
     \end{cases}
\]

where 
\begin{align}
    C(\vec{x},n_{\nu})=\frac{(E_{0,\vec{0}}-E_{1,\vec{0}})^2}{E_{0,\vec{0}}+E_{1,\vec{0}}},
\end{align}
and $\vec{x}$ is the vector describing all the Werner parameters of the initial links. If all of these states have the same Werner parameter $x$ we can rewrite $C(\vec{x},n_{\nu})$ as 
\begin{align}
C(x,n_{\nu})&=\frac{2^{n_{\nu}}x^{2n_{\nu}}}{(1 - x)^{n_{\nu}} + (1+x)^{n_{\nu}}},
\end{align}

Focusing now on the contribution of the locally sensing qubits, we will calculate $[\mathcal{F}(e^{-ih_{\nu}\phi_{\nu}}\ket{+}\bra{+}e^{ih_{\nu}\phi_{\nu}})]_{ab}$. For a pure state Eq. \ref{1parameterQFI} can be rewritten as 
 \begin{align}
     [\mathcal{F}_{\vec{\phi}}(\op{\varphi})]_{ab}&= 4\left(\langle\partial_{\phi_a}\varphi|\partial_{\phi_b}\varphi\rangle-\langle\psi|\partial_{\phi_a}\varphi\rangle\langle\partial_{\phi_b}\varphi|\varphi\rangle\right).\nonumber
 \end{align}
 For $\ket{\varphi}=e^{-ih_{\nu}\phi_{\nu}}\ket{+}_{\nu}$ , $[\mathcal{F}_{\vec{\phi}}(e^{-ih_{\nu}\phi_{\nu}}\op{+}e^{ih_{\nu}\phi_{\nu}})]_{ab}=0$ unless $a=b=\nu$. In this case,
 \begin{align}
     [\mathcal{F}_{\vec{\phi}}(e^{-ih_{\nu}\phi_{\nu}}\op{+}e^{ih_{\nu}\phi_{\nu}})]_{\nu\nu}&=4\left(\langle+|\hat{h}_\nu^2|+\rangle-|\langle+|\hat{h}_\nu|+\rangle|^2\right).
 \end{align}
 Rewriting $\ket{+}=\frac{\ket{0}+\ket{1}}{\sqrt{2}}$ and using $\hat{h}_j\ket{0}_j=\lambda_{0}\ket{0}_j$ and $\hat{h}_j\ket{1}_j=\lambda_{1}\ket{1}_j$ we get
\[   
[\mathcal{F}_{\vec{\phi}}(e^{-i\hat{h}_{\nu}\phi_{\nu}}\ket{+}\bra{+}e^{ih_{\nu}\phi_{\nu}})]_{a b} = 
     \begin{cases}
       (\Delta\lambda)^2 &\quad\text{if  } a=b=\nu\\
       \text{0} &\quad\text{otherwise,} \\ 
     \end{cases}
\]
where $
 \Delta\lambda=\lambda_{1}-\lambda_0$.
Plugging this into expression \ref{1parameterQFI} we find that 
\begin{align}
    [\mathcal{F}_{\vec{\phi}}
    (U(\vec \phi)\rho_{\vec{n}}U(\vec \phi)^{\dagger})]_{ab}=
    \begin{cases}
       C(x,n_{\nu})(\Delta\lambda)^2&\quad\text{if  } a,b\in\{1+Q_{\nu-1},\dots,Q_{\nu}\}\\
       (\Delta\lambda)^2 &\quad\text{if  } a=b\in[S-u,S]\\
       \text{0} &\quad\text{otherwise.} \\ 
     \end{cases}
\end{align}
The entries of the QFIM are nonzero if $a=b$ or if $a$ and $b$ refer to qubits in the same GHZ state. 
\subsection{QFI of $\theta$}
To get $\mathcal{F}_{\theta}$ from the QFIM matrix above, we just note that $\hat{J}=(\frac{1}{S}\ \frac{1}{S}\ \dots \frac{1}{S})^{\dagger}$ and find that
\begin{align}
    \mathcal{F}_{\theta}(\rho_{\vec{\phi}})&= \hat{J}\mathcal{F}_{\vec{\phi}}
    (U(\vec \phi)\rho_{\vec{n}}U(\vec \phi)^{\dagger})\hat{J}^{\dagger}\nonumber\\
                &= \frac{(\Delta\lambda)^2}{S^2}\left(u+\sum_{\nu=1}^{g}C(x,n_{\nu}))n_{\nu}^2\right)\nonumber\\
                &=\frac{(\Delta\lambda)^2}{S^2}\left(S+\sum_{\nu=1}^{g}\left(C(x,n_{\nu})n_{\nu}^2-n_{\nu}\right)\right),
\end{align} 
where in the last step me make use of the fact that $u=S-\sum_{\nu=1}^gn_{\nu}$.

\section{\label{Appendix:VTMBL} Average Wait-time and QFI expressions for V-TMBL Protocol}
The Variable Time Multiplexing Block Length (V-TMBL) protocol attempts entanglement generation until there are at least $\mu$ sensors sharing entangled links with the hub. Afterwards, the maximum GHZ projective measurement is performed and the entangled probe state is then used to sense. During intermediate timesteps, all the sensors will measure the desired parameter without entanglement assistance.

The probability that there are $m\geq\mu$ successes on the $T$-th timestep, but less than $\mu$ on the $(T-1)$-th timestep is given by the joint probability that $T = t$ and $M = m$ is given by:
\[    \text{Pr}(T=t,M=m)= 
    \begin{cases}
      {S\choose m}(1-p)^{S-m}p^m, & t=1\\
      {S\choose m}(1-p)^{(S-m)t}\sum_{j=m-\mu+1}^m{m\choose j}\left(1-(1-p)^{t-1}\right)^{m-j}p^j(1-p)^{j(t-1)}, & t>1,
    \end{cases}    
    \]

where $\mu\leq m\leq S$. The expected wait time then would be
\begin{align}
    \mathbb{E}[T]&={S\choose m}(\bar{p})^{S-m}p^m+{S\choose m}\sum_{j=m-\mu+1}^m{m\choose j}(\bar{p})^{S-m}p^j\sum_{t=2}^\infty t(\bar{p})^{(t-1)(S-m-j)}(1-(\bar{p})^{(t-1)})^{m-j}\nonumber\\
    &={S\choose m}(\bar{p})^{S-m}p^m+{S\choose m}\sum_{j=m-\mu+1}^m{m\choose j}(\bar{p})^{S-m}p^j\sum_{t=1}^\infty (t-1)(\bar{p})^{t(S-m-j)}(1-(\bar{p})^{t})^{m-j}\nonumber\\
    &={S\choose m}(\bar{p})^{S-m}p^m+{S\choose m}\sum_{j=m-\mu+1}^m{m\choose j}(\bar{p})^{S-m}p^j\sum_{t=1}^\infty (t-1)(\bar{p})^{t(S-m-j)}\sum_{\ell=0}^{m-j}(-1)^{\ell}{m-j\choose \ell}(\bar{p})^{\ell t}\nonumber\\
    &={S\choose m}(\bar{p})^{S-m}p^m+{S\choose m}\sum_{j=m-\mu+1}^m{m\choose j}(\bar{p})^{S-m}p^j\sum_{\ell=0}^{m-j}(-1)^{\ell}{m-j\choose \ell}(\bar{p})^{2\alpha}\left(1+(\bar{p})^{\alpha}\frac{2-(\bar{p})^{\alpha}}{(1-(\bar{p})^{\alpha})^2}\right),
\end{align}
where $\bar{p}=1-p$ and $\alpha=S-m+j+k$.

The average QFI for this protocol is 
\begin{align}
\bar{\mathcal{F}_{\theta}}&=\sum_{t=1}^{\infty}\sum_{m=\mu}^S\text{Pr}(T=t,M=m)*\left[\frac{t-1}{t}\mathcal{F}(\rho_0)+\frac{1}{t}\mathcal{F}(\rho_m)\right]\nonumber\\
&=\frac{(\Delta\lambda)^2}{S^2}\left[S+\sum_{m=\mu}^S(m^2-m)\sum_{t=1}^{\infty}\frac{\text{Pr}(T=t, M=m)}{t}\right]\nonumber\\
&=\frac{(\Delta\lambda)^2}{S^2}\left[S+\sum_{m=\mu}^S{S\choose m}(m^2-m)\left((\bar{p})^{S-m}p^m+\sum_{j=m-\mu+1}^m{m\choose j}\left(\frac{p}{\bar{p}}\right)^j\sum_{t=2}^{\infty}\frac{(\bar{p})^{t(S-m+j)}(1-(\bar{p})^{t-1})^{m-j}}{t}\right)\right]\nonumber\\
 &=\frac{(\Delta\lambda)^2}{S^2}\left[S+\sum_{m=\mu}^S{S\choose m}(m^2-m)\left((\bar{p})^{S-m}p^m+\sum_{j=m-\mu+1}^m{m\choose j}p^j(\bar{p})^{S-m}\sum_{t=1}^{\infty}\frac{(\bar{p})^{t(S-m+j)}(1-(\bar{p})^{t})^{m-j}}{t-1}\right)\right]\nonumber\\
&=\frac{(\Delta\lambda)^2}{S^2}\left[S+\sum_{m=\mu}^{S}{S\choose m}(m^2+m)\left((\bar{p})^{S-m}p^m+\sum_{j=m-\mu+1}^m{m\choose j}p^j(\bar{p})^{S-m}\sum_{\ell=0}^{m-j}(-1)^{\ell}{m-j\choose\ell}(\bar{p})^\alpha\ln{\frac{1}{1-(\bar{p})^{\alpha}}}\right)\right],
\end{align}
where again $\bar{p}=1-p$ and $\alpha=S-m+j+k$. The average latency associated with this protocol is $2L/c+(\mathbb{E}[T]-1)\tau$.

\section{\label{Appendix:Dist}Comparing Distillation Protocols}
\begin{figure}
\includegraphics[width=.8\textwidth]{ 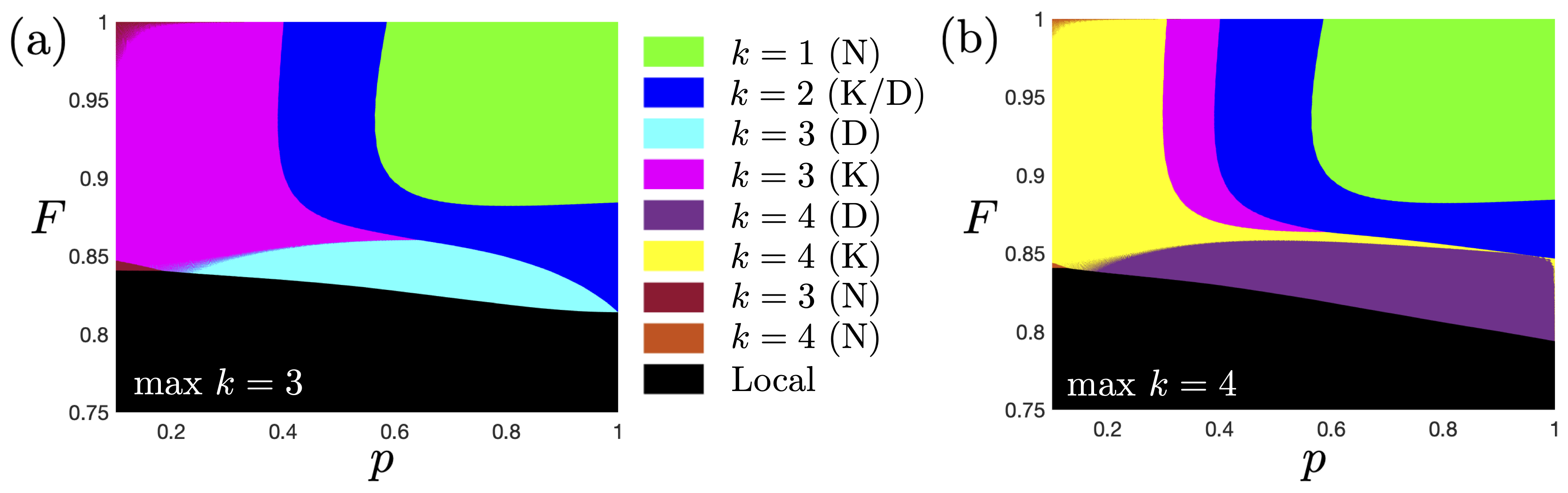}
\caption{\label{fig:9} Comparing policies for different $F$ vs $p$ and $S=5$. (a) Is for $k$ up to 3 and (b) is for $k$ up to 4. (N) - No Distillation, (D) - Distill and Discard Odd Link, (K) - Distill and Keep Odd Link }
\end{figure}
If distillation is allowed and a sensor and hub pair has $l$ successful links where $l>1$, they will repeatedly perform the 2 to 1 distillation circuit describe in the paper on states with the same fidelity. If a round of distillation fails, the sensor/hub pair might still have a lower fidelity entangled link that it can use for sensing. However, the protocol will always try and use the higher fidelity state. If no distillation attempts succeed and $l$ was even, the sensor must sense locally. If no distillation attempts succeed and $l$ was odd we can choose to \textit{keep} the remaining link with fidelity $F$ or to \textit{discard} it. The pseudocode shown in Algorithm \ref{EPSA} calculates the possible fidelities and their probability of occurring for all outcomes of the discard protocol and it can easily be updated to keep the utilize the link with fidelity $F$. In Fig. \ref{fig:9}, we compare these two distillation schemes with each other and with the no distillation option for a maximum $k$ value of three in (a) and four in (b). When $F$ is high, it is best to keep the remaining link, however when $F\sim F_{\text{thres}}(3)$ it is better to discard it. Generally, we see that distillation helps when $F<1$. However, if the maximum $k$ is fixed, there is a region in $p,\ F$ for low $p$ that not distilling links does better.

\section{\label{Appendix:SLD}Symmetric Logarithmic Derivative (SLD) Calculation}
The state after interacting with the environment can be written as
\begin{align}
    \rho_{\theta}&=\sum_{\lambda}E_{\lambda}U(\vec{\phi})\ket{\psi_{\lambda}}\bra{\psi_{\lambda}}U(\vec{\phi})^{\dagger}\nonumber\\
                &=\sum_{\lambda}E_{\lambda}\ket{e_{\lambda}(\theta)}\bra{e_{\lambda}(\theta)},
\end{align}
where $U(\vec{\phi})=e^{-i\phi^{\dagger}\hat{H}}$ and $\psi_{\lambda}$ and $E_{\lambda}$ are independent of $\theta$. The symmetric logarithmic derivative (SLD), $L$, can then be calculated as
\begin{align}
    L[\rho_{\theta}]&=2\sum_{\lambda\lambda'} \frac{\bra{e_{\lambda}(\theta)}\partial_{\theta}\rho_{\theta}\ket{e_{\lambda'}(\theta)}}{E_{\lambda}+E_{\lambda'}}\ket{e_{\lambda}(\theta)}\bra{e_{\lambda'}(\theta)}\nonumber\\
    &=2\sum_{\lambda\lambda'\lambda''} \frac{E_{\lambda''}\bra{e_{\lambda}(\theta)}\partial_{\theta} e_{\lambda''}(\theta)\rangle\bra{e_{\lambda''}(\theta)}e_{\lambda'}(\theta)\rangle+E_{\lambda''}\bra{e_{\lambda}(\theta)} e_{\lambda''}(\theta)\rangle\bra{\partial_{\theta}e_{\lambda''}(\theta)}e_{\lambda'}(\theta)\rangle}{E_{\lambda}+E_{\lambda'}}\ket{e_{\lambda}(\theta)}\bra{e_{\lambda'}(\theta)}\nonumber\\
    &=2\sum_{\lambda\lambda'} \frac{E_{\lambda'}\bra{e_{\lambda}(\theta)}\partial_{\theta} e_{\lambda'}(\theta)\rangle+E_{\lambda}\bra{\partial_{\theta}e_{\lambda}(\theta)}e_{\lambda'}(\theta)\rangle}{E_{\lambda}+E_{\lambda'}}\ket{e_{\lambda}(\theta)}\bra{e_{\lambda'}(\theta)}\nonumber\\
    &=2\sum_{\lambda\lambda'}\frac{E_{\lambda'}\bra{e_{\lambda}(\theta)}\partial_{\theta}e_{\lambda'}(\theta)\rangle+E_{\lambda}\bra{\partial_{\theta}e_{\lambda}(\theta)}e_{\lambda'}(\theta)\rangle}{E_{\lambda}+E_{\lambda'}}\ket{e_{\lambda}(\theta)}\bra{e_{\lambda'}(\theta)}\nonumber\\
    &=2\sum_{\lambda\lambda'}\frac{E_{\lambda'}-E_{\lambda}}{{E_{\lambda}+E_{\lambda'}}} \bra{e_{\lambda}(\theta)}\partial_{\theta} e\rangle\ket{e_{\lambda}(\theta)}\bra{e_{\lambda'}(\theta)},\nonumber\\
\end{align}
where we use the fact $\partial_{\theta}(\bra{e_{\lambda}(\theta)}e_{\lambda'}(\theta)\rangle)=0$ so $\bra{\partial_{\theta}e_{\lambda}(\theta)}e_{\lambda'}(\theta)\rangle=-\bra{e_{\lambda}(\theta)}\partial_{\theta}e_{\lambda'}(\theta)\rangle$  if $\lambda\neq\lambda'$ for the last simplification. Taking the partial derivative we get $\partial_{\theta}U(\vec{\phi})\ket{\psi_{\lambda}}=S^{-1}\sum_{s=1}^S\partial_{\phi_s}U(\vec{\phi}\ket{\psi_{\lambda}})=-iS^{-1}\sum_{s=1}^Sh_sU(\vec{\phi})\ket{\psi_{\lambda}}$ so that,
\begin{align}\label{SLDeq}
     L[\rho_{\theta}]&=\frac{2i}{S}\sum_{\lambda\lambda'}\frac{E_{\lambda}-E_{\lambda'}}{{E_{\lambda'}+E_{\lambda}}}\bra{\psi_{\lambda}(\theta)}\sum_{s=1}^Sh_s\ket{\psi_{\lambda'}(\theta)}\ket{e_{\lambda}(\theta)}\bra{e_{\lambda'}(\theta)}.
\end{align}
Our probe state consist of $g$ different GHZ diagonal states -- corresponding to the $g-$disjoint GHZ measurements performed at the hub. Let the $\nu$-th GHZ state be $n_{\nu}$-GHZ diagonal. Lets define $u=S-\sum_{\nu=1}^{g}n_{\nu}$ to be the total number of sensors using local probe states. The probe state then can be written as:
\begin{align}
    \rho_{\vec{n}}=\left(\bigotimes_{\nu=1}^{g}\sigma_{n_{\nu}}\right)\otimes\ket{+}\bra{+}^{\otimes u},
\end{align}
where $\vec{n}=(n_1\ n_2\ \cdots\ n_g)$ and $\sigma_{n_{\nu}}$ is a GHZ diagonal state whose explicit form can be found in Appendix \ref{Appendix:GHZcoeff}. Continuing to use the notation from Appendix \ref{Appendix:GHZcoeff}, we can express $\rho_{\vec{n}}$ in terms of the eigenvectors $\ket{\Psi_{\vec{j}\ \vec{i}_1\cdots\vec{i}_g}}$, defined as 
\begin{align}
\ket{\Psi_{\vec{j}\ \vec{i}_1\cdots\vec{i}_g,\ \vec{\ell}}}=\ket{\psi_{j_1,\vec{i}_1}\cdots\psi_{j_g,\vec{i}_g}}\bigotimes_{\nu=S-u}^S\frac{\ket{0}_{\nu}+(-1)^{\ell_{\nu}}\ket{1}_{\nu}}{\sqrt{2}},
\end{align}
where $\vec{j}=(j_1,\dots,j_{g})^{\dagger}$ and $\vec{\ell}$ is a binary vector string of length $u$ denoting the eigenvectors of the unentangled qubits. If $\ell_{\nu}=0$,  the $\nu$-th eigenvector is  $\ket{+}_{\nu}$ , whereas if  $\ell_{\nu}=1$, the $\nu$-th eigenvector is in state $\ket{-}_{\nu}$. With this notation, the probe state can be expressed as
\begin{align}
\rho_{\vec{n}}=\sum_{\vec{j},\ \vec{i_1}\dots\vec{i_g}}\xi_{\vec{j}\ \vec{i}_1\cdots\vec{i}_g\ \vec{0}}\ket{\Psi_{\vec{j}\ \vec{i}_1\cdots\vec{i}_g\ \vec{0}}}\bra{\Psi_{\vec{j}\ \vec{i}_1\cdots\vec{i}_g\ \vec{0}}},
\end{align}
where 
\begin{align}
    \xi_{\vec{j}\ \vec{i}_1\cdots\vec{i}_g\ \vec{\ell}}=\delta_{\vec{\ell}\vec{0}}\prod_{a=1}^gE_{j_a\vec{i}_a}.
\end{align}

We can rewrite Eq \ref{SLDeq} by expressing $\hat{h}_s$ in terms of $Z_s$, where $Z_s$ applies a $Z$-gate to the $s$th qubit and applies an identity to all others. Then $\hat{h}_s=-\frac{\Delta\lambda}{2}\hat{Z}_s+(\lambda_{0}+\frac{\Delta\lambda}{2})\hat{\mathbb{I}}$, which satisfies $\hat{h}_s\ket{0}=\lambda_{0}\ket{0}_s$ and $\hat{h}_s\ket{1}=\lambda_{1}\ket{1}_s$ where $\Delta\lambda=\lambda_{1}-\lambda_{0}$.

The second term, $(\lambda_{0}+\frac{\Delta\lambda}{2})\bra{\Psi_{\vec{j}'\ \vec{i}_1'\cdots\vec{i}_g'\vec{\ell}}}\Psi_{\vec{j}\ \vec{i}_1\cdots\vec{i}_g\vec{\ell}}\rangle$, is only
 nonzero iff $\vec{j}=\vec{j}'$, $\vec{i}_s=\vec{i}_s' \forall s \in \{1,\cdots,g\}$, $\vec{\ell}=\vec{\ell}'$. However, if this is the case then $\xi_{\vec{j}'\ \vec{i}'_1\cdots\vec{i}'_g\ \vec{\ell}'}-\xi_{\vec{j}\ \vec{i}_1\cdots\vec{i}_g\ \vec{\ell}}=0$, so the coefficient in the front of Eq. \ref{SLDeq} will be zero.
 
The first term is only nonzero when $\ket{\Psi_{\vec{j}'\ \vec{i}_1'\cdots\vec{i}_g'\ \vec{\ell}'}}=\hat{Z}_s\ket{\Psi_{\vec{j}\ \vec{i}_1\cdots\vec{i}_g\ \vec{\ell}}}$ since 
\begin{align}
-\frac{\Delta\lambda}{2}\bra{\Psi_{\vec{j}'\ \vec{i}_1'\cdots\vec{i}_g'\ \vec{\ell}'}}\hat{Z}_s\ket{\Psi_{\vec{j}\ \vec{i}_1\cdots\vec{i}_g\ \vec{\ell}}}=-\frac{\Delta\lambda}{2}\bra{\Psi_{\vec{j}\ \vec{i}_1\cdots\vec{i}_g\ \vec{\ell}}}\hat{Z}_s\hat{Z}_s\ket{\Psi_{\vec{j}\ \vec{i}_1\cdots\vec{i}_g\ \vec{\ell}}}\nonumber=-\frac{\Delta\lambda}{2}\bra{\Psi_{\vec{j}\ \vec{i}_1\cdots\vec{i}_g\ \vec{\ell}}}\hat{\mathbb{I}}\ket{\Psi_{\vec{j}\ \vec{i}_1\cdots\vec{i}_g\ \vec{\ell}}}=-\frac{\Delta\lambda}{2}.
\end{align}
Therefore the only nonzero entries are when $\vec{i}=\vec{i}'$ and either one entry of $\vec{j}$ differs from  $\vec{j'}$ strictly or one entry of $\vec{\ell}$ differs from  $\vec{\ell'}$. When $s$ and $s'$ both refer to qubits $\nu$-th GHZ diagonal state, it doesn't matter which qubit the $Z$ gate is applied to so $Z_{s}\ket{\Psi_{\vec{j}\ \vec{i}_1\cdots\vec{i}_g\vec{\ell}}}=Z_{s'}\ket{\Psi_{\vec{j}\ \vec{i}_1\cdots\vec{i}_g\vec{\ell}}}$. Then
\begin{align}
    \bra{\Psi_{\vec{j'}\ \vec{i'}_1\cdots\vec{i'}_g \vec{\ell}'}}\sum_{s=1}^SZ_s\ket{\Psi_{\vec{j}\ \vec{i}_1\cdots\vec{i}_g \vec{\ell}}}=
    \left\{
    \begin{array}{lr}
        n_{\nu}, & \text{if } j_{\nu}\neq j'_{\nu}, j_{s}=j_s'\ \forall s \in\{1,\cdots, g\}/\nu,\vec{i}_s=\vec{i'}_s \forall s \in\{1,\cdots, g\}, \vec{\ell}=\vec{\ell}'\ \\
         1, & \text{if } \vec{j}=\vec{j}', \vec{i}_s=\vec{i'}_s \forall s \in\{1,\cdots, g\}, \ell_s=\ell'_s\ \forall s \in\{1,\cdots, g\}/a, \ell_a\neq\ell_a' \\
        0 & \text{else. }
    \end{array}
\right\}
\end{align}
Now, updating our notation, lets define $Z_{\nu}$ to be a $Z$ gate applied to any qubit in the $\nu$-th GHZ state. Also, if  $\xi_{\vec{j},\vec{i}_1\dots\vec{i}_g\ \vec{\ell}}$ is the eigenvalue of the state $\Psi_{\vec{j},\vec{i}_1\dots\vec{i}_g\ \vec{\ell}}$ then let $\xi^{(Z_\nu)}_{\vec{j},\vec{i}_1\dots\vec{i}_g\ \vec{\ell}}$ denote the eigenvalue of the state $Z_{\nu}\Psi_{\vec{j},\vec{i}_1\dots\vec{i}_g\ \vec{\ell}}$. Then we can write the SLD as 
\begin{align}
    L[\rho_{\theta}]=\frac{2i}{S}\sum_{\nu=1}^{g+u}n_{\nu}\sum_{\vec{j}\vec{i}_1\cdots\vec{i}_g\ \vec{\ell}}\frac{\xi^{(Z_\nu)}_{\vec{j},\vec{i}_1\dots\vec{i}_g\ \vec{\ell}}-\xi_{\vec{j},\vec{i}_1\dots\vec{i}_g\ \vec{\ell}}}{\xi^{(Z_\nu)}_{\vec{j},\vec{i}_1\dots\vec{i}_g\ \vec{\ell}}+\xi_{\vec{j},\vec{i}_1\dots\vec{i}_g\ \vec{\ell}}}U(\vec{\phi})\ket{\Psi_{\vec{j},\vec{i}_1\dots\vec{i}_g\ \vec{\ell}}}\bra{\Psi_{\vec{j},\vec{i}_1\dots\vec{i}_g\ \vec{\ell}}}Z_{\nu}U(\vec{\phi})^{\dagger},
\end{align}
where $n_{\nu}$ is equal to the size of the $\nu$-th GHZ projective measurement and $n_{\nu}=1$ for all $\nu\in[g+1,g+u]$.

 If we want to explicitly separate out the local sensors, then we get 
\begin{align}
    L[\rho_{\theta}]=\frac{2i}{S}U(\vec{\phi})\bigg[\sum_{\nu=1}^{g}n_{\nu}\sum_{\vec{j}\ \vec{i}_1\cdots\vec{i}_g ;\vec{i}_{\nu}=\vec{0}}\frac{E^{(Z)}_{{j}_{\nu},\vec{i}_{\nu}}-E_{{j}_{\nu},\vec{i}_{\nu}}}{E^{(Z)}_{{j}\ \vec{i}_{\nu}}+E_{{j}_{\nu},\vec{i}_{\nu}}}\ket{\Psi_{\vec{j},\vec{i}_1\dots\vec{i}_g\ \vec{0}}}\bra{\Psi_{\vec{j}_{},\vec{i}_1\dots\vec{i}_g\ \vec{0}}}Z_{\nu}\nonumber\\+\sum_{\nu=S-u}^{\nu=S}\left(\ket{-}\bra{+}-\ket{+}\bra{-}\right)\otimes\pi_{/\nu}\bigg]U(\vec{\phi})^{\dagger},
\end{align}
where $\pi_{/\nu}$ is the maximally mixed state for all the qubits not in the $\nu$-th state. In this step, we have also plugged in for $\xi_{\vec{j}\ \vec{i}_1\cdots\vec{i}_g\ \vec{\ell}}$ . Further manipulation gets us

\begin{align}
    L[\rho_{\theta}]=\frac{2i}{S}U(\vec{\phi})\bigg[\sum_{\nu=1}^{g}n_{\nu}\frac{E_{1,\vec{0}}-E_{0,\vec{0}}}{E_{1,\vec{0}}+E_{0,\vec{0}}}\left(\ket{\psi_{0,\vec{0}}}_{\nu}\bra{\psi_{1,\vec{0}}}-\ket{\psi_{1,\vec{0}}}_{\nu}\bra{\psi_{0,\vec{0}}}\right)\otimes\pi_{/\nu }\nonumber\\+\sum_{\nu=S-u}^S\left(\ket{-}_{\nu}\bra{+}-\ket{+}_{\nu}\bra{-}\right)\otimes{\pi_{/\nu}}\bigg]U(\vec{\phi})^{\dagger}.
\end{align}

If we diagonalize the $L[\rho_{\theta}]$, the eigenvectors are
\begin{align}
U(\vec{\phi})\bigotimes_{\nu=1}^{g+u}\ket{c(n_{\nu},d)} \text{       where }\ket{c(n_{\nu},d)}=\frac{1+(-1)^d i}{2}\ket{0}^{\otimes n_{\nu}}+\frac{1-(-1)^di}{2}\ket{1}^{\otimes n_{\nu}},
\end{align}
and $d\in\{0,1\}$. In this equation, $n_{\nu}$ is equal to the size of the $\nu$-th GHZ projective measurement and $n_{\nu}=1$ for all $\nu\in[g+1, \dots, g+u]$, along with the set of vectors orthogonal to these states. Together, these are a set of optimal projective measurements to perform when measuring $\theta$. 

\section{\label{Appendix:LocalMeasurement}Local Measurement}
Let there be $S$ Werner states, where the $s$-th state can be written $\beta_{A_sB_s} = x_S \Phi_{A_sB_s} + \frac{(1-x_s)}{4}\mathbb{I}_{A_s}\otimes\mathbb{I}_{B_s}$ where $\Phi_{A_sB_s}$ corresponds to a Bell pair between two qubits $A_s$ and $B_s$. The Werner parameter of this state is $x_s=\frac{4F_s-1}{3}$ where $F_s$ is the fidelity. For now lets assume that all the Werner states have the same initial fidelity $F$ and therefore the same Werner parameter $x$. One qubit of each pair will be measured off to prepare the state:
\begin{equation}
 \rho_{S}=\bra{\Gamma(S)}_{B_1\cdots B_{S}}\bigotimes_{s=1}^{S}\beta_{A_sB_s}\ket{\Gamma(S)}_{B_1\cdots B_{S}}.
\end{equation}
The exact for of $\rho_S$ can be found in Appendix \ref{Appendix:GHZcoeff}.
After interacting with the environment, the state becomes $\rho_{\theta}=U(\vec{\phi})\rho_SU(\vec{\phi})^{\dagger}$. One local measurement strategy we could use is to measure every qubit locally in the $\ket{+},\ \ket{-}$ basis. This set of POVMs can be written as
\begin{align}
\Pi_{c_1\cdots c_S}=\bigotimes_{y=1}^S\left(\ket{0}_y+(-1)^{c_y}\ket{1}_y\right)\left(\bra{0}_y+(-1)^{c_y}\bra{1}_y\right) \ \ \ \      \text{where } \ket{c_y}\in \{0,1\}.
\end{align} 
The $y$-th qubit being measured in the $\ket{+}$ state corresponds to $c_y=0$ while a measurement in the $\ket{-}$ state corresponds to $c_y=1$. 
The probability of measuring all the qubits in the $\ket{+}$ is 
\begin{align}
    \text{Pr}(c_1 \cdots c_S)&=\text{Tr}[\Pi_{c_1\cdots c_S}U(\vec{\phi})\rho_{S}U(\vec{\phi})^{\dagger}]\nonumber\\
    &=\frac{1+(-1)^{\sum_{y=1}^Sc_y}x^S\cos(\Delta\lambda \sum_{j=1}^S\phi_j)}{2^{S}},
\end{align}
From this we can calculate the classical Fisher information:
\begin{align}
    \text{CFI}(S,x)&=\sum_{c_1\cdots c_S}\frac{\left(\frac{\partial}{\partial\theta}\text{Pr}(c_1\cdots c_S)\right)^2}{\text{Pr}(c_1\cdots c_S)},
\end{align}
where $\frac{\partial}{\partial\theta}\text{Pr}(c_1\cdots c_S)=\frac{1}{S}\{\frac{\partial}{\partial\phi_1}+\cdots+\frac{\partial}{\partial\phi_S}\}\text{Pr}(c_1\cdots c_S)$. Plugging in, we get

\begin{align}
    \text{CFI}(S,x)&=\frac{(\Delta\lambda)^2}{2S^2}\left[ \frac{x^{2S}\sin^2(\Delta\lambda\sum_{j=1}^S\phi_j)}{1+x^S\cos(\Delta\lambda \sum_{j=1}^S\phi_j)}+\frac{x^{2S}\sin^2(\Delta\lambda\sum_{j=1}^S\phi_j)}{1-x^S\cos(\Delta\lambda \sum_{j=1}^S\phi_j)}\right]\nonumber\\
    &=\frac{(\Delta\lambda)^2}{S^2}\frac{x^{2S}\sin^2(\Delta\lambda\sum_{j=1}^S\phi_j)}{1-x^{2S}\cos^2(\Delta\lambda\sum_{j=1}^S\phi_j)}\nonumber\\
    &=\frac{(\Delta\lambda)^2}{S^2}\frac{x^{2S}\sin^2(S\Delta\lambda \theta)}{1-x^{2S}\cos^2(S\Delta\lambda\theta)},
\end{align}
where in the last set we substituted $\theta=\frac{\phi_1+\cdots+\phi_S}{S}$. If the $S$ Werner states have different initial Werner parameters s.t. the $s$-th state has Werner parameter $x_{s}$, it is trivial to show that the above expression becomes
\begin{align}
    \text{CFI}(S,x_1\cdots x_S)=\frac{(\Delta\lambda)^2}{S^2}\frac{\prod_{s=1}^Sx_{s}^2\sin^2(S\Delta\lambda \theta)}{1-\prod_{s=1}^Sx_{s}^2\cos^2(S\Delta\lambda\theta)}.
\end{align} 

Now, lets look at $u$ local probes states (un-entangled) prepared in in the $\ket{+}$ state and measured in the $\ket{+},\ \ket{-}$ basis.  It is trivial to show that this is a QFI achieving measurement for local probes, meaning the CFI is equal to $(\Delta\lambda)^2u/S^2$.

Using the additivity of classical Fisher information \cite{Frieden_2010} we can combine the above results to get a more general expression for CFI. Let the probe state have $g$ different GHZ diagonal states where the $\nu$-th state is $n_{\nu}$-GHZ diagonal. This means that we have $u=S-\sum_{\nu=1}^gn_{\nu}$ unentangled probes.

We can order the qubits such that the  $n_1$ qubits in the $n_1$-GHZ diagonal state are at sensors $\{1,\cdots,n_1\}$ and the  $n_2$ qubits in the $n_2$-GHZ diagonal state are at sensors $\{n_1+1,\cdots,n_1+n_2\}$  and so forth. To refer to qubits in a certain GHZ-diagonal state, let 
\begin{align}
    Q_{\nu}=\sum_{i=1}^{\nu-1}n_i,
\end{align}
where we will define $Q_1=0$. The qubits at sensors $\{S-u,\cdots,S\}$ are all local probes.
\begin{align}
    \text{CFI}(\vec{n},x_1=\cdots= x_{S-u})=\frac{(\Delta\lambda)^2}{S^2}\left(u+\sum_{\nu=1}^g\frac{\sin^2(\Delta\lambda\sum_{q=1+Q_{\nu}}^{n_{\nu}+Q_{\nu}}\phi_q)\prod_{s=1+Q_{\nu}}^{s=n_{\nu}+Q_{\nu}}x_{s}^2}{1-\cos^2(\Delta\lambda \sum_{q=1+Q_{\nu}}^{n_{\nu}+Q_{\nu}}\phi_q)\prod_{s=1+Q_{\nu}}^{s=n_{\nu}+Q_{\nu}}x_{s}^2}n_{\nu}^2\right).
\end{align}

% Produces the bibliography via BibTeX.
\end{document}